\documentclass{elsart}
\usepackage{tabularx}
\usepackage[hang]{caption2}
\usepackage{longtable}
\usepackage{lscape}
\usepackage{rotating}
\usepackage{natbib}
\usepackage{graphics}
\usepackage{graphicx}
\usepackage{epsfig}
\usepackage{amssymb}
\usepackage{thumbpdf}
\usepackage{lineno}

\begin{document}
\begin{frontmatter}
 \title{Mechanical Spectroscopy on Volcanic Glasses}
 \thanks[talk]{norman.wagner@mfpa.de}
\author{Norman Wagner}
\address{Institute of Material Research and Testing  at the Bauhaus-University Weimar,\\ Amalienstr. 13, 99423 Weimar
        \\Tel: ++49-3643-564-221\\Fax: ++49-3643-564-202}
\author{Klaus Heide}
\address{Institute of Geoscience, Friedrich-Schiller-University Jena,\\ Burgweg 11, 07743 Jena}

\begin{abstract}
Mechanical relaxation behaviour of various natural volcanic glasses
 have been investigated in the temperature range
$RT-1000^\circ$C using special low frequency flexure (f$\approx$0.63Hz) pendulum experiments. The
rheological properties complex Young's modulus $M^\star(\omega,\tau)$ and internal friction
$Q^{-1}(\omega, \tau)$ have been studied from a pure elastic solid at room temperature to pure
viscous melt at $\log(\eta[Pas])=8$. The Young's modulus at room temperature $M_{RT}=(70\pm10)$GPa
is nearly constant. There is a positive correlation with the water content and a weak negative
correlation with the cooling rate. Several relaxation processes are assumed to act: the primary
$\alpha$-relaxation (viscoelastic process, $E_a=(344...554)kJ/mol$) above the glass transition
temperature $T_g=(935...1105)K$ and secondary anelastic $\beta',\beta$ and $\gamma$-relaxation
processes below $T_g$. The dynamic glass transition, i.e. the viscoelastic $\alpha$-transition,
can be characterized with hierarchically coupled relaxation processes which lead to an equivalent
distribution of relaxation times nearly independent of the fragility for all examined glasses. The
observed secondary relaxation processes can be explained with different mechanisms: ($\gamma,
\beta$) cooperative movement of alkali ions in the vitreous state. ($\beta'$) cooperative movement
of alkaline earth ions and non bridging oxygen's near the glass transition range. Here the
influence of water must be taken into account as well as alteration effects due to structural
$\alpha$-relaxation. With a simple fractional Maxwell model with asymmetrical relaxation time
distribution, $H(\tau)$, phenomenological the mechanical relaxation behaviour, is described. This
establish a basis of realistic concepts for modelling of volcanic or magmatic processes.
\end{abstract}
\end{frontmatter}
%\tableofcontents
%\begin{linenumbers}
%--------------------------------------%--------------------------------------
\section{Introduction}\label{sec:Introduction}
%--------------------------------------%--------------------------------------

\begin{figure}[ht]
  \begin{center}
    \includegraphics[scale=0.6]{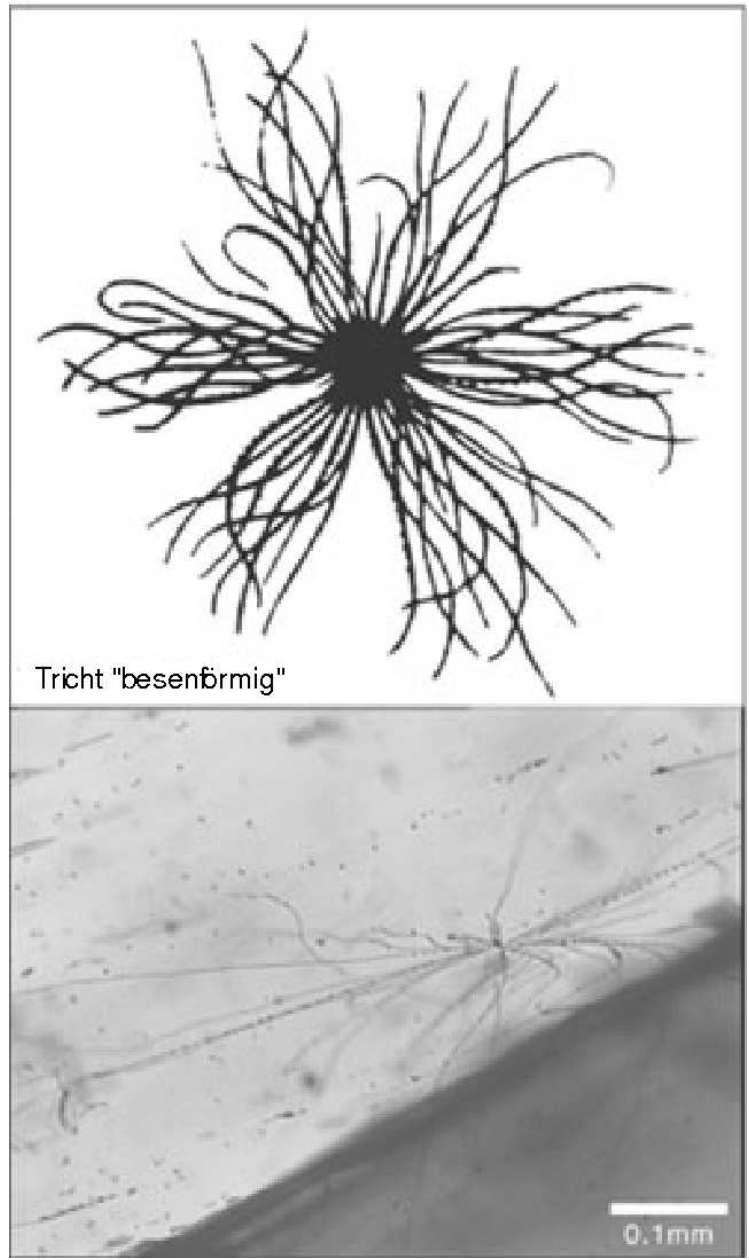}
    \caption{(top) Copperplate engraving by Zirkel 1873 of a ``magnetite spider'' in an  obsidian from Armenia
         (adapted from \cite{Zirk73}). (bottom) Turbulent  hairline magnetite (fibres) in vitreous matrix of obsidian from Graber,
            Armenia together with fluidal adjusted bubbles.}
   \label{fig:Zirkel}
  \end{center}
\end{figure}

It is of great importance today, in material and geoscience, to be able to understand and predict
the mechanical response (elasticity, anelasticity and viscoelasticity) of multicomponent silicate
glasses and melts(\cite{Pink92}, \cite{Webb97}, \cite{Bark97}, \cite{Donth01}, \cite{Schil03},
\cite{Duan03}, \cite{Muel03}, \cite{Buch03}, \cite{Wagn04c}). The evolution of terrestrial planets
is determined by the rheological properties of silicate melts and their stability under different
atmospheric conditions. An important roll plays the water content as well as the chemical
composition of the natural materials which reflect and cause the evolution processes
(\cite{Meln99}). By optical microscopy we know already since more than one hundred years textural
patterns in vitreous and crystalline volcanic rocks (\cite{Zirk73}, \ref{fig:Zirkel}).
%%% ------------------------------------
%%% ------------------------------------
This observations clearly indicate degassing induced crystallization as well as a high mobility of
crystalline phases in the magma and lava during the emplacement process (\cite{Heid89},
\cite{Spark00}). According to theoretical and experimental investigations of \cite{Gutz96}
formation of such patterns are only possible below viscosities $\log(\eta [Pas])$=4, i.e. at
temperatures far above the glass transformation range. In the case of rhyolitic glasses at
approximately $1300^{\circ}C$ under atmospheric conditions, i.e. far above field experience of
emplacement of dacitic lava at $\approx 925^{\circ}C$ (Tab. \ref{tab:temp}).

\begin{table}[b]
  \centering
  \caption{Estimated extrusion temperatures from field observations as well as calculations from the mineralogy
           of various high viscose lavas (\cite{Hall96})}\label{tab:temp}
  \begin{tabular}{|l|l|l|}
    \hline
    Locality & Rock type & Temperature [$^\circ C$]\\\hline
    %Kileauea, Hawaii & Tholeitic basalt & 1150-1225\\
    Taupo, New Zealand & Rhyolite lava and pumice & 735-890\\
    Mono Craters, USA & Rhyolite lava & 790-820\\
    Island & Rhyodacite obsidian & 900-925\\
    New Britain & Andesite pumice & 940-990\\
    New Britain & Dacite lava and pumice & 925\\
    New Britain & Rhyodacite pumice & 880\\\hline
  \end{tabular}
\end{table}

In addition in daily life silicate melts products are ubiquitous. Natural handling with glass
usually misleads us to believe that the glass transition is one of the ten great outstanding
challenges in physics (\cite{Gutz01}, \cite{Ngai00}, \cite{Ange00}, \cite{Donth01},
\cite{Debe01}).

From a geoscientific point of view, experimental vulcanology and petrology examine the structure,
dynamics and properties of natural and synthetic silicate systems (\cite{Perch91}, \cite{Saxe92},
\cite{Steb95}, \cite{Myse03}). The rheological properties determine emplacement mechanisms
generally and the hazard implications especially (\cite{Fink90}, \cite{Smit97}, \cite{Meln99}).
Whereby fragmentation of vesicular magma is one of the main processes governing explosive
volcanism (\cite{Ding96}, \cite{Papa99}, \cite{Gonn03}). Besides the material consuming and
dangerous direct observation of lava flows it is also in some cases possible to reconstruct the
emplacement history from microscopic texture in combination with experimental techniques under
laboratory conditions (\cite{Heid89}, \cite{Fink90}, \cite{Mang98}, \cite{Cast99}, \cite{Mang01},
\cite{Rust02}, \cite{Cast02}, \cite{Cano04}, \cite{Cano04a}). As products of very viscous melts
volcanic (rhyolitic in particular) glasses (obsidian) are the link to the melt. The
characterization of rheological properties of these volcanic glasses are the subject of this work.

The material property, by which glass formation is determined, is the viscosity $\eta$ or the
structural or $\alpha$- relaxation time $\tau_{\alpha}$. Their dependence on pressure $P$,
temperature $T$, chemical composition $X$, bubble content $c_b$, crystal content $c_{xx}$ and
volatile content $c_v$ represents the goal of intensive geoscientific investigations
(\cite{Bott72}, \cite{Shaw72}, \cite{Rich84}, \cite{Humm85}, \cite{Perc91},
 \cite{Pink92}, \cite{Stei92},  \cite{Bagd94}, \cite{Stev95}, \cite{Leje95},
\cite{Stev96}, \cite{Bake96}, \cite{Hess96},
\cite{Bagd97},   \cite{Mang98},  \cite{Stev98},
 \cite{Bagd00}, \cite{Mang01}, \cite{Saar01}, \cite{Sipp01},  \cite{Gior03}).

Since in the glass transition range the relaxation time reaches orders of magnitude, which are no
longer accessible under laboratory conditions. Thus, the transition from viscous melt to solid
elastic body, i.e. the viscoelastic response, determines the material behavior  (\cite{Vers92},
\cite{Bagd93}, \cite{Bagd93a}, \cite{Bagd97}, \cite{Bagd99}, \cite{Bagd01}, \cite{Webb97},
\cite{Duff97}, \cite{Duff97a}, \cite{Duff97b}, \cite{Duff97c}, \cite{Duff98}, \cite{Muel03},
\cite{Wagn03}, \cite{Webb03}).

In the vitreous state the structure of the melt is frozen in, however, here in addition dynamic
processes (mobility of ions in the glassy network) take place, which can be measured by mechanical
spectroscopy as internal friction, whereby a small water content cause unresolved phenomena
(\cite{Roet41}, \cite{Fitz51}, \cite{Fitz51a}, \cite{Roet57}, \cite{Roet58}, \cite{Ryde61},
\cite{Coen61}, \cite{Day62}, \cite{Day62a}, \cite{Day69}, \cite{Shel69}, \cite{Mcva70},
\cite{Shel70}, \cite{Day73}, \cite{Roet74}, \cite{Roet75}, \cite{Day74}, \cite{Day74b},
\cite{Day74a}, \cite{VanA74}, \cite{Tayl74}, \cite{Zdan76}, \cite{Tayl79a}, \cite{Tayl79b},
\cite{Bart83}, \cite{Vers92}, \cite{Ke96}, \cite{Bart96}, \cite{Zoll98}, \cite{Roli98},
\cite{Roli98a}, \cite{Roli01}).

In this study the rheological properties, Young's moduls $M^\star$ and internal friction
$\tan\delta=Q^{-1}$, of natural volcanic glasses have been investigated in a wide temperature
range from a purely elastic solid at room temperature to purely viscous melt at a viscosity
$\log(\eta [Pas])\approx 8$. It will contribute to the understanding of the thermocinetics and
dynamics that govern the glass transition, at temperatures well above, below and at the glass
transition temperature $T_g$ (\cite{Donth01}). In this context the rheological properties of
natural and synthetic silicate glasses and melts are of interest. With a simple generalized
fractional Maxwell model, the relaxation behavior is described phenomenologically. This establish
a basis of realistic concepts for modelling of volcanic or magmatic processes.

%------------------------------------------------------------------------
\section{Basics}
%------------------------------------------------------------------------

\begin{figure}[ht]
  \begin{center}
    \includegraphics[scale=0.55]{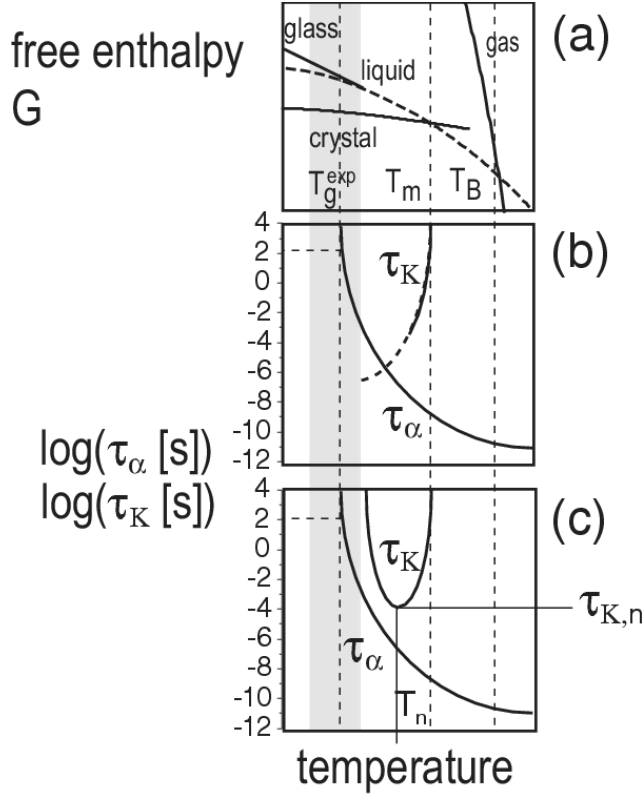}
    \caption{Schematic representation to the construction of the $TTT$-curve (time-temperature-transformation).
    (a) isobar dependence of the free enthalpy $G$ on temperature $T$ of a gas, a liquid (stable and supercooled liquid),
    and a crystal as well as the corresponding glass of a pure substance with the boiling point $T_B$, melting point $T_m$
    and the glass transition temperature $T_g$.
    (b) The hypothetical intersection point between $\tau_K$ and $\tau_\alpha$ is not reached, since (c)
    $\tau_K$ goes up strongly due to the drastic viscosity increase with $\tau_\alpha$ (after \cite{Debe96}).}
    \label{fig:TTT}
  \end{center}
\end{figure}

A necessary and sufficient condition to glass formation is a corresponding (critical) cooling rate
$q_K$ to avoid cristallization. A coarse approximation for the estimate of $q_K$ gives
\cite{Owen85}, \cite{Feltz83} and \cite{Debe96}:
%++++++++++++++++++++++++++++++++
\begin{equation}
  \label{eq:kritCool}
  q_K=2\cdot10^{-6}\frac{T_m^2R}{V\eta(T_n)}=\frac{T_m-T_n}{\tau_{K,n}}
\end{equation}
%++++++++++++++++++++++++++++++++
with the fusing temperature $T_m$, the gas constant $R$, the molar volume $V$, viscosity $\eta$,
the so called nose temperature $T_n$ and corresponding critical time $\tau_{K,n}$. Here, according
to \cite{Debe96} or \cite{Ange88} the time $\tau_K$ for formation of a critical volume $V_K$ as
well as the structural relaxation time $\tau_\alpha$ and their interplay
(time-temperature-transformation) is of great significance (cf. Fig. \ref{fig:TTT}).
%%% ------------------------------------
%%% ------------------------------------
Above the thermal glass transition temperature $T_{g}^{exp}$ specified by the experimental
conditions, which is determined by a characteristic structural or $\alpha$ relaxation time
$\tau_{\alpha}$, the mechanical glass transition occurs at the `mechanical' relaxation time
$\tau_{\alpha}(T_\alpha)\omega\approx1$. The mechanical glass transition temperature $T_{\alpha}$
shifts with measurement frequency $\omega=2\pi f$ in general via the Arrhenian equation
%-------------------------------
\begin{equation}
  \label{eq:Arrhenius}
  \ln\left(\frac{\Omega}{\omega}\right)=\frac{E_a(T)}{k_BT_\alpha}
\end{equation}
%-------------------------------
with jump frequency $\log(\Omega [Hz])\approx12-14$
as typical frequency of molecular vibration (\cite{Russ03}), Boltzmann-constant $k_B$ and a
temperature dependent activation energy $E_a(T)$ with $\frac{dE_a}{dT}\leq 0$ (\cite{Gutz95},
\cite{Bart96}). The most common
entrances to the modelling of the temperature dependence of the activation energies are:

\newcounter{Fakt0a}
\begin{list}{(\textbf{\arabic{Fakt0a}})}{\usecounter{Fakt0a}}
\item the empirical VFT-equation (\cite{Voge21}, \cite{Fulc25}, \cite{Tamm26}) with the adjustable
parameters $B_{VFT}$ and $T_0$ ($T_0$ is usually equated with the Kauzmann-temperature
(\cite{Hodg94}, \cite{Hodg97}))
%%% ------------------------------------
\begin{equation}
  \label{eq:VFT-Aktivierung}
  E_a(T)=\frac{B_{VFT}}{1-\frac{T_0}{T}}.%=\frac{B_{VFT}T}{T-T_0}
\end{equation}
%%% -----------------------------------
\item the Adam-Gibbs-equation  (\cite{Adam65}) with the adjustable parameter $B_{AG}$ and a
temperature dependend configurational entropy $S_c(T)$
%%% ------------------------------------
\begin{equation}
  \label{eq:AG-Aktivierung1}
  E_a(T)=\frac{k_BB_{AG}}{S_c(T)}.
\end{equation}
%%% -----------------------------------
\item the Avramov-equation (\cite{Avra88}) with a dimensionless activation energy $\varepsilon_A$,
the Avramov fragility index $\alpha_A$  and the glass transition temperature $T_g$ at a viscosity
of $\log(\eta [Pas])=12.3$
%%% ------------------------------------
\begin{equation}
   \label{eq:Avramov-Aktivierung-Temperatur}
   E_a(T)=\frac{k_B\varepsilon_A T_g^{\alpha_A}}{T^{\alpha_A-1}}.
\end{equation}
%%% -----------------------------------
\item the Williams-Landel-Ferry-equation (WLF, (\cite{Will55}) with the parameters $A_{WLF}$,
$B_{WLF}$ and the fragility index $m=\frac{d\ln(\tau_\alpha)}{d(T_g/T)}\left|_{T_g}\right.$ at the
glass transition temperature $T_g$ (\cite{Boeh92}, \cite{Donth01})
%%% ------------------------------------
\begin{equation}
   \label{eq:WLF}
   E_a(T)=mT-\frac{A_{WLF}T(T-T_g)}{B_{WLF}+(T-T_g)}.
\end{equation}
%%% -----------------------------------
\end{list}

$\log(\Omega)$ and the Kauzmann-Temperature
$T_0\approx(48...815)K$ are the asymptotes concerning the
frequency or temperature (\cite{Dont81}, \cite{Donth92},
\cite{Donth01}, \cite{Hodg94}, \cite{Ange00}).

The mechanical glass transition for small deformations $10^{-4}$ is the linear viscoelastic
response of the material. Then the linear response is completely determined by properties of the
equilibrium melt if thermal fluctuations are included in the equilibrium concept. The
fluctuation-dissipation theorem (FDT) of statistical physics implies the important statement that
it is exclusively thermal fluctuations which determine the linear response (\cite{Donth01},
\cite{Roli01}). Especially for the understanding of processes\footnote{Within glass science these
processes called 'Thermometereffekt' according to internal friction investigations of
\cite{Roet41}.} on a geological time scale these concepts are very useful (\cite{Nemi03}).
Nonlinearity has to be take into account for large deformations and high strain rates, i.e. under
emplacement conditions (\cite{Bagd93}, \cite{Drag94}, \cite{Drag96}, \cite{Renn00}, \cite{Blak00},
\cite{Buis02}).

Above a critical temperature $T_C$ predicted by the idealized mode coopling theory, $MCT$, the
system is retained to ergoic and below non-ergodic (\cite{Goet92}). Non-ergodicity means that the
correlation function $\Phi$ does not converge against an expected equilibrium value of
$\lim_{t\rightarrow\infty}\Phi(t)=\Phi_\infty=0$, but reaches a final value $\Phi_\infty>0$. Since
the corresponding coupling parameters are temperature dependent, one finds the critical
temperature $T_C$ if the existence of a non-ergodic stage to the first time have been observed. In
this sense one speaks of a phase transition at the glass transition but The precise designation is
egodicity transition. That is, that density fluctuations froze below $T_C$. Above $T_C$ the
structural $a$-relaxation and second $\beta$-relaxation can not be distinguished. Below $T_C$
diffusion processes and the structural $\alpha$-relaxation are to be observed separately. In the
internal friction spectrum the characteristic secondary $\beta$ loss maxima occur. Ergodicity can
be restored by introduction of hopping-processes in the enlarged MCT also below $T_C$. The
temperature dependence of the $a$-time scale for $T>T_C$ is of the form (\cite{Schn00},
\cite{Donth01}):
%%% ------------------------------------
\begin{equation}
  \label{eq:MCTAlpha}
  \tau_a\propto\eta\propto(T-T_C)^\gamma.
\end{equation}
%%% ------------------------------------
with a parameter $\gamma$. The following general rule is valid (\cite{Buch03a})
%%% ------------------------------------
\begin{equation}
  \label{eq:MCTAlpha1}
  \tau_{\alpha}(T_c)\approx10^{-7}s.
\end{equation}
%%% ------------------------------------
The distinction of the dynamics of the primary $a$ process and a $\beta_{fast}$ process (in the
sense of a microscopically fast movement) above $T_C$ in the field of the undercooled liquid is
predicted, however, through the $MCT$ and confirmed by neutron and light scattering experiments
onto different glass formers as well as through computer simulations (\cite{Roes96a},
\cite{Siew97}, \cite{Horb02}, \cite{Buch03a}). The main problem with the $MCT$ consists in the
analysis of the $\alpha$-process. So (\ref{eq:MCTAlpha}) can only be observed experimentally in a
small temperature range. If one identifies $T_C$ with the Vogel temperature $T_0$ or the glass
transition temperature $T_g$, then experimental relaxation times and (\ref{eq:MCTAlpha}) are not
to bring into agreement. A comparison with experimental data leads to the reasonable statement
$T_C>T_g>T_0$ and $T_C/T_g\approx1.3...2.6\propto 1/m$ for silicate melts (\cite{Hess96a},
\cite{Pfei98}). There according to the mode coupling theory for $T<T_C$ the $\alpha$ process no
more is supposed to occur but is experimentally even strongly observed at $T\approx T_g$. For this
reason one is today generally of the opinion, that the mode coupling approximation is a good
theory for $T > T_C$ and break down for temperatures below $T_C$ especially for strong melts. In
spite of these difficulties the $MCT$ is today the widest driven forward analytical
representation, that understands essential appearances of the glass transition correctly and
contains also many possibilities of a generalization (\cite{Schu00}).
%%% ------------------------------------
\begin{figure}[ht]
  \begin{center}
    \includegraphics[scale=0.5]{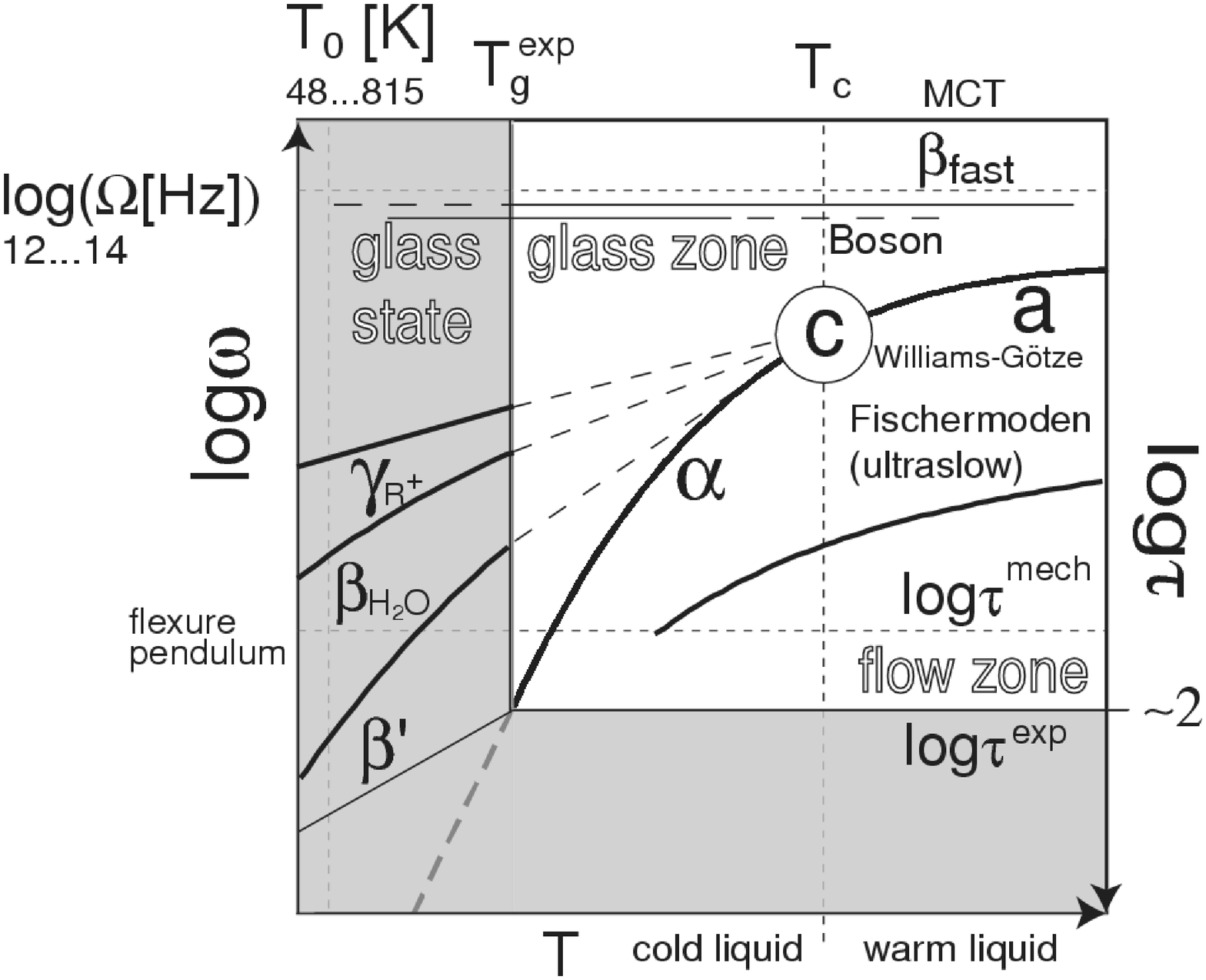}
    \caption{Schematic representation of thermocinetic and dynamic $\alpha$-transition of an inorganic-glass or melt and related phenomena:
$\gamma_{R^+}$-alkali ion transport, $\beta_{H_2O}$ - water peak, $\beta'$- superposition of
Johari-Goldstein relaxation and the thermal glass transition (c.f. \cite{Doli02}).}
   \label{fig:RelaxMapNaSi}
  \end{center}
\end{figure}
%%% ------------------------------------

During cooling, the mobility of the atoms decrease.  At the (nonlinear) thermal glass transition
range (the relaxation time is a function of temperature $T$ and the structural state
$\tau(T,T_f)$) typical relaxation times are reached by seconds or minutes, in order to take the
(metastable) equilibrium position. If the temperature now falls, the structure can not reach their
equilibrium and is therefore ``trapped" in a definite configurational state at the limit of the
fictive temperature $T_f$ (\cite{Tool46}, \cite{Gard70}, \cite{Mart85}). \cite{Ritl56},
\cite{Mart87}, \cite{Rich92}, \cite{Wild95}, \cite{Wild96}, \cite{Wild96a}, \cite{Wild00},
\cite{Gott01}, \cite{Gott01a},  \cite{Debe01}, \cite{Tang01}). In the vitreous state it is
possible to measure $\beta$ or secondary mechanical relaxation processes which are substantially
decoupled of the structural alpha relaxation (c.f. Fig. \ref{fig:RelaxMapNaSi}). This processes
can be explained with different mechanisms: ($\gamma, \beta$) cooperative movement of alkali ions
in the vitreous state. ($\beta'$) cooperative movement of alkaline earth ions and non bridging
oxygen's near the glass transition range. Here the influence of water must be taken into account
as well as alteration effects due to structural $\alpha$-relaxation. The temperature dependents of
diffusion, electrical conductivity and mechanical relaxation processes can be described in most
cases by an Arrhenian-law via equation (\ref{eq:Arrhenius}) with an equivalent activation energy
(see \cite{Fris75}, \cite{Paul98}, \cite{Maas99}, \cite{Ngai00}).

Relaxation processes (ion mobility, viscoelasticity) deviating
from the classical exponential (Debye) behavior are often
encountered in the dynamics of complex materials such as silicate
glasses below and above $T_g$, glass ceramics and partially molten
rocks (\cite{Jons77}, \cite{Muel83}, \cite{Kamp85}, \cite{Bagd93},
\cite{Webb97},  \cite{Roli98}, \cite{Ange00}, \cite{Donth01},
\cite{Ngai00}). In many cases experimentally observed relaxation
functions exhibit a stretched exponential
(Kohlrausch-Williams-Watts) decay \cite{Kohl47}
%----------------------
\begin{equation}
  \label{eq:KWW}
  \Phi\propto \exp\left(-\left[\frac{t}{\tau}\right]^{\beta_{KWW}}\right)
\end{equation}
%----------------------
with $0<\beta_{KWW}<1$, or a scaling decay
%----------------------
\begin{equation}
  \label{eq:Power}
  \Phi\propto \left[\frac{t}{\tau}\right]^a
\end{equation}
%----------------------
with $0<a<1$ (\cite{Bagd93}, \cite{Bagd93a}, \cite{Bagd97}, \cite{Bagd99}, \cite{Bagd00},
\cite{Muel03}). An appropriate tool to describe phenomenologically this richness of dynamical
features is fractional calculus ($FC$, see below) (\cite{Schi95}).

%------------------------------------------------------------------------
\subsection{Rheological models}
%------------------------------------------------------------------------

Usually phenomenologic viscoelastic models are based on springs and dampers. A spring is purely
elastic and describes the instantaneously acting Hookc law with the dynamic responses (complex
modulus $M^\star=M'+iM''$ with storage modulus $M'$ and loss modulus $M''$ or complex creep
compliance $M^\star\cdot D^\star=1$) being constant and real (M''/M'=$Q^{-1}=0$). The constitutive
equation of stress $\sigma$ and strain $\varepsilon$ is of the form
%----------------------
\begin{equation}
  \label{eq:Hooke}
  \sigma(t)=M_0\varepsilon(t),\hspace{0.5cm}M^\star=M'=M_0=D_0^{-1}
\end{equation}
%----------------------
with an idealized Young's modulus $M_0$. A damper is purely viscous and describes the
instantaneously acting Newton law with purely imaginary dynamic responses ($Q^{-1}=\infty$)
\cite{Donth92}:
%----------------------
\begin{equation}
  \label{eq:Newton}
  \sigma(t)=\eta\frac{d\varepsilon(t)}{dt},\hspace{0.5cm}M^\star=iM''=i\omega\eta
\end{equation}
%----------------------
Through combinations of springs and dashpots one arrives at standard viscoelastic models, such as
the Maxwell or the Zener, these models involve a fairly small number of single elements. The
problem here is that the corresponding ordinary differential equations have a relatively
restricted class of solutions, which is, in general, too limited to provide an adequate
description for the complex systems discussed in the introduction. To overcome this shortcoming
one can
 relate stress and strain through fractional equations (\cite{Schi95}):
%----------------------
\begin{equation}
  \label{eq:Fractional}
  \sigma(t)=M_0\tau^\beta\frac{d^\beta\varepsilon(t)}{dt^\beta}
\end{equation}
%----------------------
with $0\leq\beta\leq1$  and the relaxation time $\tau$. In this way one readily obtains scaling decays.
In general,$FC$ allows the interpolation between the purely elastic behaviour of equation (\ref{eq:Hooke}),
obtained for $\beta=0$ in equation (\ref{eq:Fractional}), and the purely viscous pattern
of equation (\ref{eq:Newton}), obtained for  $\beta=1$  in equation (\ref{eq:Fractional}).
\cite{Schi93, Schi95, Schi95a, Schi95b, Heym94, Heym96} and \cite{Heym96} have
demonstrated that the fractional relation, equation (\ref{eq:Fractional}), can be realized
physically through hierarchical arrangements of springs and dampers, such as ladders,
 trees or fractal structures. We now introduce the term fractional element ($FE$) to denote such a
hierarchical structure and specify it by the trible $(\beta, M_0, \tau)$.

%--------------------------------------%--------------------------------------
\subsection{Fractional Maxwell model}
%--------------------------------------%--------------------------------------

The classical viscoelastic Maxwell model (composed of a linear elastic element and a
linear viscous element in series) may be modified by replacing the damper by a $FE$.
The dynamic responses for this model is:
%----------------------
\begin{equation}
  \label{eq:Caputo}
   \frac{M^\star(\omega, \tau)}{M_0}=\frac{(i\omega\tau)^\gamma}{1+(i\omega\tau)^\gamma},\hspace{0.5cm}
   \frac{D^\star(\omega, \tau)}{D_0}=1+(i\omega\tau)^{-\gamma}
\end{equation}
%----------------------
which is the so called Caputo-Model (\cite{Bagd00}, \cite{Bagd99}) or Cole-Cole-Model
(\cite{Cole41}). In the high temperature and low frequency limit (\ref{eq:Caputo}) leads to
constant-$Q^{-1}$. For  $\gamma=1$ (\ref{eq:Caputo})  is transformed to the classical Maxwell
model. When both elements are replaced by $FE$'s the dynamic responses is of the Jonscher-type
(\cite{Jons77}):
%----------------------
\begin{equation}
  \label{eq:FracMaxwell}
   \frac{M^\star(\omega, \tau)}{M_0}=\frac{(i\omega\tau)^\alpha}{1+(i\omega\tau)^{\alpha-\beta}},\hspace{0.5cm}
   \frac{D^\star(\omega, \tau)}{D_0}=(i\omega\tau)^{-\alpha}+(i\omega\tau)^{-\beta}
\end{equation}
%----------------------
with $\beta<\alpha$ and the constitutive equation is of the form:
%----------------------
\begin{equation}
  \label{eq:FractionalConst}
  \sigma(t)+\tau^{\alpha-\beta}\frac{d^{\alpha-\beta}\sigma(t)}{dt^{\alpha-\beta}}=
   M_0\tau^\alpha\frac{d^\alpha\varepsilon(t)}{dt^\alpha}
\end{equation}
%----------------------
Since $0\leq\beta<\alpha\leq1$ the condition for thermodynamic compatibility is fulfilled
(\cite{Heym94}, \cite{Heym96}). Transformation into a standard model is carried out by adding an
elastic element of modulus $M_1$ in series. The dynamic response $M^\star(\omega, \tau)$  and
internal friction $Q^{-1}$ then become:
%----------------------
\begin{eqnarray}
  \label{eq:FracMaxwell1}
   \frac{M^\star(\omega, \tau)}{M_0}=\frac{M_1(i\omega\tau)^\alpha}
         {M_1(1+(i\omega\tau)^{\alpha-\beta})+M_0(i\omega\tau)^\alpha},\\\nonumber\\
   \frac{D^\star(\omega, \tau)-D_1}{D_0}=(i\omega\tau)^{-\alpha}+(i\omega\tau)^{-\beta}\nonumber
\end{eqnarray}
%----------------------
%----------------------
\begin{equation}
  \label{eq:q-FracMaxwell1}
   Q^{-1}(\omega, \tau)=\frac{(\omega\tau)^{-\alpha}\sin\left(\frac{\alpha\pi}{2}\right)+
                      (\omega\tau)^{-\beta}\sin\left(\frac{\beta\pi}{2}\right)}
                     {(\omega\tau)^{-\alpha}\cos\left(\frac{\alpha\pi}{2}\right)+
                      (\omega\tau)^{-\beta}\cos\left(\frac{\beta\pi}{2}\right)+\frac{M_0}{M_1}}
\end{equation}
%----------------------
It is also possible to give an analytical expression for the relaxation time distribution
$h(\tau)$ with the relation $M^\star(\omega, \tau)=\hat{M}(p,\tau)|_{p=i\omega}$ \cite{Gloe93}:
%----------------------
\begin{equation}
  \label{eq:Distribution}
   h(\tau)=\pm\lim_{\varepsilon\rightarrow 0}\Im[\hat{M}(p)]|_{p=-1/\tau\pm\varepsilon}
\end{equation}
%----------------------
and one gets:
%----------------------
\begin{equation}
  \label{eq:Distribution1}
   h(\tau)=\frac{\Omega_{\alpha, \beta}^S+\frac{M_0}{M_1}}
              {\tilde{\tau}^{-2\alpha}+\tilde{\tau}^{-2\beta}+
              2\tilde{\tau}^{-\alpha-\beta}\cos([\alpha-\beta]\pi)+
              \frac{2M_0}{M_1}\Omega_{\alpha, \beta}^{C}+\frac{M_0^2}{M_1^2}}
\end{equation}
%----------------------
with $\tilde{\tau}=\tau/\tau_{max}$,
$\Omega_{i,j}^{C}=\tilde{\tau}^{-i}\cos(i\pi)+\tilde{\tau}^{-j}\cos(j\pi)$,
$\Omega_{i,j}^{S}=\tilde{\tau}^{-i}\sin(i\pi)+\tilde{\tau}^{-j}\sin(j\pi)$.
%\newpage
\begin{landscape}
%\begin{sidewaystable}
\begin{table}[p]
%  \centering
  \caption{The summary of various spectral functions and their power-law behaviour.
   Abbreviation: $p=i\omega\tau$, $A',A$ - constants.}
  \label{tab:Jonscher}
  \begin{tabular}{|l|l|l|l|l|l||l|l|l|}
    \hline
   %      \multicolumn{9}{|c|}{} \\
   \multicolumn{3}{|c|}{}& \multicolumn{3}{c|}{${\omega\tau\rightarrow0}$} & \multicolumn{3}{c|}{${\omega\tau\rightarrow\infty}$} \\
   \hline\hline
    %     \multicolumn{9}{|c|}{} \\\hline
     & $M^\star/M_0$ & &     $M'/M_0$ & $M''/M_0$ & $Q^{-1}$ & $M'/M_0$ & $M''/M_0$ & $Q^{-1}$ \\
    \hline
    %**************************************************************************************************
    \cite{Deby41} & $(1+p^{-1})^{-1}$ &
    & $(\omega\tau)^2$ & $(\omega\tau)^1$ & $(\omega\tau)^{-1}$ & 1 & $(\omega\tau)^{-1}$ & $(\omega\tau)^{-1}$ \\
    %**************************************************************************************************
   \cite{Cole41}  & $(1+p^{-\alpha})^{-1}$              & $0<\alpha\leq1$ & $(\omega\tau)^\alpha$ & $(\omega\tau)^\alpha$
   & $A'$ & 1 & $(\omega\tau)^{-\alpha}$ & $(\omega\tau)^{-\alpha}$ \\
    %**************************************************************************************************
    \cite{Jons77}    & $(p^{-\alpha}+p^{-\beta})^{-1}$     & $0<\beta\leq\alpha\leq1$ & $(\omega\tau)^\alpha$
   & $(\omega\tau)^\alpha$ & $A'$ & $(\omega\tau)^\beta$ & $(\omega\tau)^{\beta}$ & $A$ \\
    %**************************************************************************************************
    KWW (\cite{Kohl47})   & $\exp(-(t/\tau)^\beta)$             & $0<\beta\leq1$ & $(\omega\tau)^{2}$ & $(\omega\tau)^1$
   & $(\omega\tau)^{-1}$ & 1 & $(\omega\tau)^{-\beta}$ & $(\omega\tau)^{-\beta}$ \\
    %**************************************************************************************************
    \cite{Havr67}    & $1-(1+p^{\alpha})^{-\beta}$ & $0<\alpha,\beta<1$& $(\omega\tau)^{\alpha}$
   & $(\omega\tau)^{\alpha}$ & $A'$ & 1 & $(\omega\tau)^{-\beta\alpha}$ & $(\omega\tau)^{-\beta\alpha}$ \\
    %**************************************************************************************************
    \cite{Davi51}   & $1-(1+p)^{-\beta}$   & $0<\beta\leq1$ & $(\omega\tau)^{2}$ & $(\omega\tau)^{1}$
   & $(\omega\tau)^{-1}$  & 1 & $(\omega\tau)^{-\beta}$ & $(\omega\tau)^{-\beta}$ \\
    %**************************************************************************************************
    \cite{Muel83}   & $Q=(\omega\tau)^\gamma$   & $0<\gamma\leq1$ & $\rightarrow0$ & $(\omega\tau)^{1}$
   & $(\omega\tau)^{-\gamma}$  & 1 & $(\omega\tau)^{-\gamma}$ & $(\omega\tau)^{-\gamma}$ \\\hline
    %**************************************************************************************************
    FC1   & $(p^{-1}+p^{-\beta}+\frac{M_0}{M_1})^{-1}$        & $0<\beta\leq1$ & $(\omega\tau)^{2-\beta}$ & $(\omega\tau)^1$
   & $(\omega\tau)^{-1+\beta}$ & 1 & $(\omega\tau)^{-\beta}$ & $(\omega\tau)^{-\beta}$ \\
    %**************************************************************************************************
    FC2   & $(p^{-\alpha}+p^{-\beta}+\frac{M_0}{M_1})^{-1}$   & $0<\beta\leq\alpha\leq1$ & $(\omega\tau)^{\alpha}$
   & $(\omega\tau)^\alpha$ & $A'$ & 1 & $(\omega\tau)^{-\beta}$ & $(\omega\tau)^{-\beta}$ \\
    \hline
  \end{tabular}
\end{table}

%\end{sidewaystable}
\end{landscape}
\newpage
%-----------------------------------
\section{Sample selection}

%----------------------%----------------------
\begin{table}\label{tab:chemiche_zusammensetzung}
  \begin{center}
    \caption{Chemical composition and water-content  (wt.\%) of the volcanic glasses.
               \small{$<$dtl: less then dedection limit, agpaitic index (mol\%)
             $AI=\frac{K_2O+Na_2O}{Al_2O_3}$, aluninium satturation index
             $ASI=\frac{Al_2O_3}{K_2O+Na_2O+CaO}$, structur modifier (mol\%) $SM$
              from \cite{Gior03},
             exess oxides (mol\%) $EO=SM-0.5Fe_2O_{3t}-Al_2O_3$
             from \cite{Gott02}}.}
    \begin{tabular}{|llllllllll|}
      \hline
  &  YEL  &  VUL  &  LIP  &  MIL  &  RAB  &  DYR  &  ATS  &  JAL  &  QUI  \\\hline
$SiO_2$  &  77.01  &  74.66  &  74.84  &  77.03  &  74.56  &  78.67  &  78.57  &  76.39  &  77.11  \\
$TiO_2$  &  $<$dtl &  $<$dtl &  $<$dtl &  0.22  &  $<$dtl &  0.26  &  $<$dtl &  $<$dtl &  0.13  \\
$Al_2O_3$  &  11.97  &  12.84  &  12.92  &  12.72  &  13.04  &  11.6  &  11.95  &  10.79  &  14.06  \\
$Fe_2O_{3t}$  &  1.53  &  1.98  &  1.86  &  1.34  &  1.88  &  0.6  &  0.57  &  3.53  &  0.64  \\
$CaO$  &  0.54  &  0.81  &  0.86  &  1.52  &  0.89  &  1.6  &  1.34  &  0.18  &  0.46  \\
$MgO$  &  $<$dtl &  $<$dtl &  $<$dtl &  $<$dtl &  $<$dtl &  $<$dtl &  $<$dtl &  $<$dtl &  $<$dtl \\
$MnO$  &  $<$dtl &  $<$dtl &  $<$dtl &  $<$dtl &  $<$dtl &  $<$dtl &  $<$dtl &  $<$dtl &  $<$dtl \\
$Na_2O$  &  3.11  &  3.29  &  3.87  &  3.05  &  4.04  &  0.7  &  2.31  &  3.94  &  2.55  \\
$K_2O$  &  5.61  &  5.92  &  5.2  &  3.98  &  5.17  &  6.28  &  5.07  &  5.05  &  4.56  \\\hline
Total  &  99.77  &  99.5  &  99.88  &  99.86  &  99.94  &  99.71  &  99.81  &  99.88  &  99.51
\\\hline\hline
AI$^{-1}$  &  1.07  &  1.08  &  1.08  &  1.36  &  1.06  &  1.46  &  1.29  &  0.92  &  1.55  \\
ASI  &  0.99  &  0.96  &  0.95  &  1.05  &  0.94  &  1.07  &  1.02  &  0.88  &  1.42  \\
$\frac{NBO}{T}$  &  0.0103  &  0.0197  &  0.0257  &  2E-4  &  0.0199  &  0.0037  &  0.005  &  0.023  &  0.409  \\
EO  &  0.47  &  0.89  &  1.16  &  -0.17  &  0.9  &  -0.38  &  0.23  &  1.05  &  -1.97  \\
H$_2$O &  0.096  &  0.163  &  0.21  &  0.113  &  0.101  &  0.085  &  0.106  &  0.027  &  0.222  \\
\hline
    \end{tabular}
  \end{center}
\end{table}
%%% ------------------------------------

The nine natural volcanic glasses used were fresh, unweathered, unaltered, and nonhydrated
obsidians free of cracks with low crystal ($<1\%$) and bubble content ($<1\%$).
Chemical composition of the matrix glass was quantified by electron microanalysis (SEM-EDX,
DSM 940 CARL ZEISS, eXL 10 Spektrometer, Oxford Instruments).
Volatile- species and thus water content are determined in high-vacuum degassing experiments
(controlled heating rate 10K/min from RT-1500C) coupled with a quadropole mass spectrometer
(QMA 125 Balzers, \cite{Schm00}, \cite{Lesc03}, \cite{Heid03a}).

$YEL$ is a grey crystal-poor obsidian from Yellowstone (USA).
$VUL$ from Vulcano, $LIP$ from Lipari (Italy) and $JAL$ from Jalisco (Mexico) are grey crystal-free glasses.
The sample $MIL$ from Milos (Greek) is grey-black crystal- poor. $RAB$ from Hrafntinnuhryggur
(Island) and $DYR$ from Artenis (Armenia) are black crystal-poor glasses. The obsidian $ATS$ from Artis
(Armenia) is grey crystal- poor with crystal bands. $QUI$ from Quironcolo (Argentinia) and $IKI$ from Ikizidre
(Turkey) are transparent homogeneous obsidians with very few large crystals.

%--------------------------------------%--------------------------------------
\section{Experiments}
%--------------------------------------%--------------------------------------

%%% ------------------------------------
\begin{figure}[h]
  \begin{center}
    \includegraphics[scale=0.55]{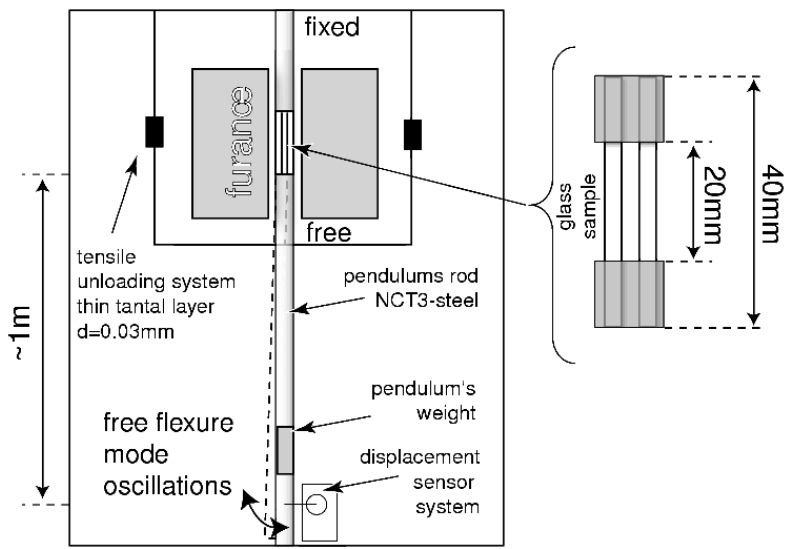}
    \caption{Flexure Pendulum Equipment.}
    \label{fig:Pendulum}
  \end{center}
\end{figure}
%%% ------------------------------------

The flexure pendulum is shown in Fig. \ref{fig:Pendulum}. It operates in air at a frequency of
about $0.63Hz$ (small deformations $\approx 10^{-5}...10^{-3}$) and temperatures up to $T\approx
1000^\circ C$ (cf. \cite{Bark94}, \cite{Bark97a}, \cite{Bark97}, \cite{Bark98a},  \cite{Zoll98},
\cite{Bark98b}, \cite{Wagn01}). The sample is a combination of two bars with a rectangular cross
section of (1x1)mm$^2$ and a lenght $l$ of 40mm. One end of the specimen is held rigid and the
other end is connected with the movable part of the flexure pendulum. Measurements were done under
isothermal conditions in the range below the glass transition temperature $T_g$ in 10K-steps and
above $T_g$ in 5K-steps. Experimentally determined quantities are the logarithmic decrement
without specimen $\Lambda_{P}$ and of the coupled system specimen-pendulum $\Lambda_{SP}$,
oscillation period of pendulum $t_P=f_P=2\pi\omega_P$ and of the coupled system specimen-pendulum
$t_{SP}$ . With the direction moment $D_P=(5.35\pm0.36)Nm$ of the pendulum and geometrical
quantities of the specimen: free length $l$, moment of inertia $I$ and experimentally determined
quantities rheological parameters complex Young's modulus $M^\star(\omega_P, T)$:
%----------------------
\begin{equation}
  \label{eq:PendelReal}
  M'(\omega_P, T)=\frac{3}{2}\frac{D_Pl}{I}\left(\frac{t_P^2}{t_{SP}^2}-1\right),
\end{equation}
\begin{equation}
  \label{eq:PendelImag}
  M''(\omega_P, T)=\frac{3}{2}\frac{D_Pl}{\pi I}\left(\Lambda_{SP}-\Lambda_P\right)
\end{equation}
%----------------------
and internal friction $Q^{-1}(\omega_P, T)$ were calculated
\begin{equation}
  \label{eq:PendelTanDelta}
  Q^{-1}(\omega_P, T)=\frac{M''(\omega_P, T)}{M'(\omega_P, T)}=
                            \frac{t_{SP}^2(\Lambda_{SP}-\Lambda_P)}{\pi(t_P^2-t_{SP}^2)}.
\end{equation}
%----------------------
%%% ------------------------------------
\begin{figure}[p]
  \begin{center}
    \includegraphics[scale=0.71]{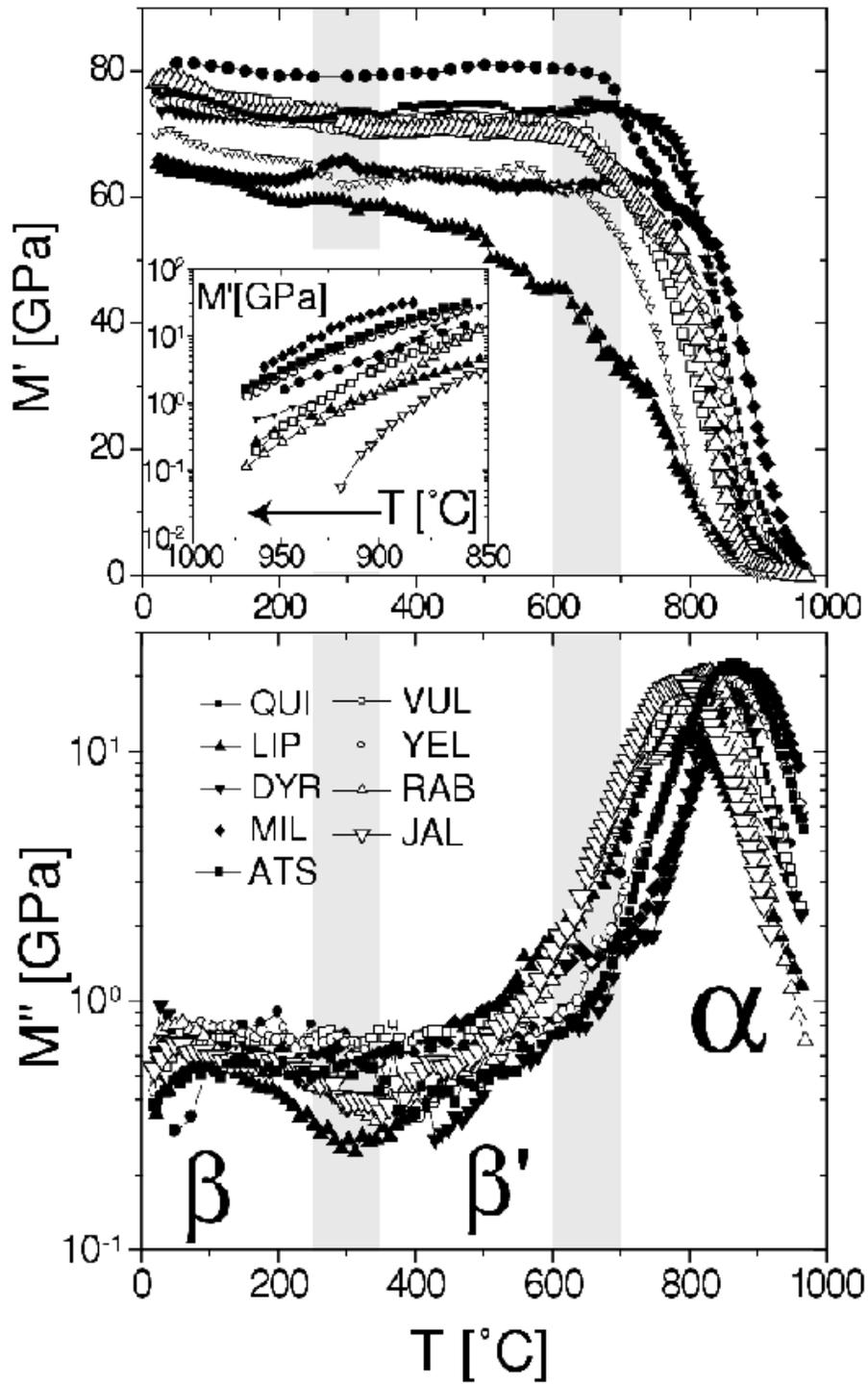}
 \caption{Temperature dependence of the real part (storage modulus) $M'$ and imaginary part (loss modulus)
                $M''$of the complex Young's modulus
                 $M^\star=M'+iM''$ for the natural volcanic glasses. (inset) Asymptotic behaviour of storage modulus at hight
                temperature.}
%    \caption{Temperature dependence of the imaginary part (loss modulus) $M''$ of the complex Young's modulus
%                 $M^\star$ for the natural volcanic glasses.}
   \label{fig:Module}
  \end{center}
\end{figure}
%%% ------------------------------------

The mechanical spectra are characterized by the complex Young's
modulus $M^\star(T)$, complex shear modulus $G^\star(T)$, internal
friction $Q^{-1}(T)$, complex creep compliance
$D^\star(T)M^\star(T)=1$, shear compliance
$J^\star(T)G^\star(T)=1$ and/or the complex shear viscosity
$\eta^\star(t)$.  The complex quantities can be separated into
real (storage modulus $M'$) and imaginary parts (loss modulus
$M''$). The Kramers-Kronig (dispersion) relations (\cite{Kram26},
\cite{Kron26}) couple the real and imaginary parts of the complex
quantities $A^\star=A'+iA''$ of a material by
%----------------------
\begin{equation}
  \label{eq:linearResponse4}
   A'(\omega)=-H[A''(\xi)], \hspace{1cm}A''(\omega)=H[A'(\xi)]
\end{equation}
%----------------------
with the Hilbert-transform by principal value integral $\oint$ (\cite{Donth01})
%----------------------
\begin{equation}
  \label{eq:Hilbert}
   H[f(\xi)](\omega)=\frac{1}{\pi}\oint\frac{f(\xi)d\xi}{\xi-\omega},\hspace{1cm}H^{-1}=-H.
\end{equation}
%----------------------
These are purely mathematical implications of the linear and causal material equations.
The discussion of the mechanical spectra can take place thereby on the individual modulus.

%%% ------------------------------------
\begin{figure}[htp]
  \begin{center}
    \includegraphics[scale=0.5]{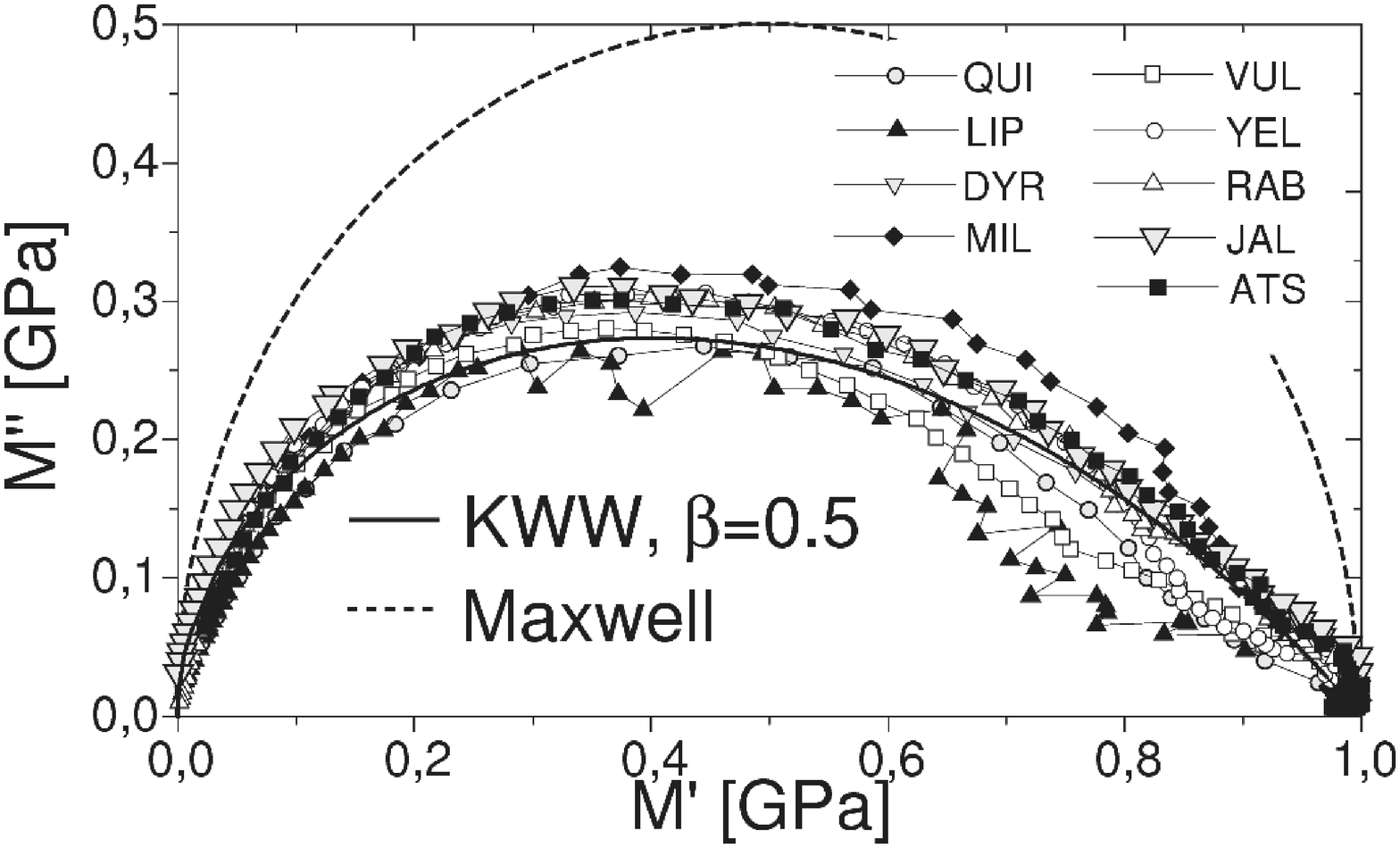}
    \caption{Complex Youngs modulus $M^\star$ normalised
          with the unrelaxed Youngs modulus $M_{\alpha}$ at $T_g$ in the Gaussian-plane.}
   \label{fig:Argand}
  \end{center}
\end{figure}
%%% ------------------------------------

%--------------------------------------%--------------------------------------
\section{Results}\label{sec:results}
%--------------------------------------%--------------------------------------

Fig.\ref{fig:Module} represents the storage modulus $M'$ and the loss modulus $M''$ for the
investigated natural glasses. All obsidians show a relatively similar behaviour.  A deviation from
the general trend shows the $LIP$ sample due to strong vesiculation.  Here, a particularly
long-drawn-out thermal and mechanical glass transition can be observed in the storage modulus.

\begin{figure}[p]
  \begin{center}
    \includegraphics[scale=0.5]{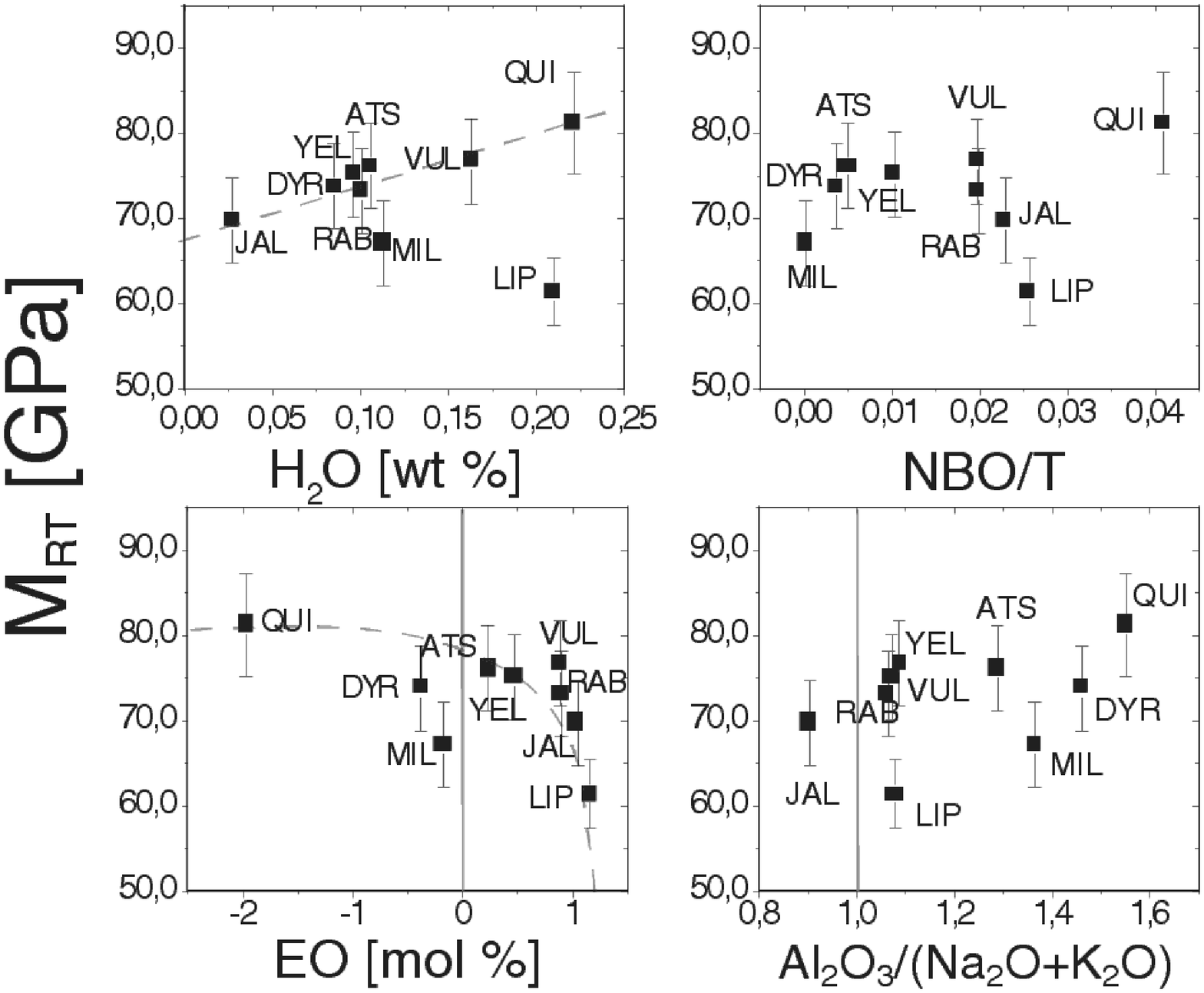}
    \caption{Dependence of the Young's modulus $M_{RT}$ on the chemical composition
    (according to $NBO/T$ as well as $EO$ and $AI^{-1}$ to Tab. \ref{tab:chemiche_zusammensetzung})
    as well as the water content for the examined obsidians. Lines are guides to the eye.}
    \label{fig:WasserModul}
%  \end{center}
%\end{figure}
%%% ------------------------------------
%%% ------------------------------------
%\begin{figure}[h]
%  \begin{center}
    \includegraphics[scale=0.4]{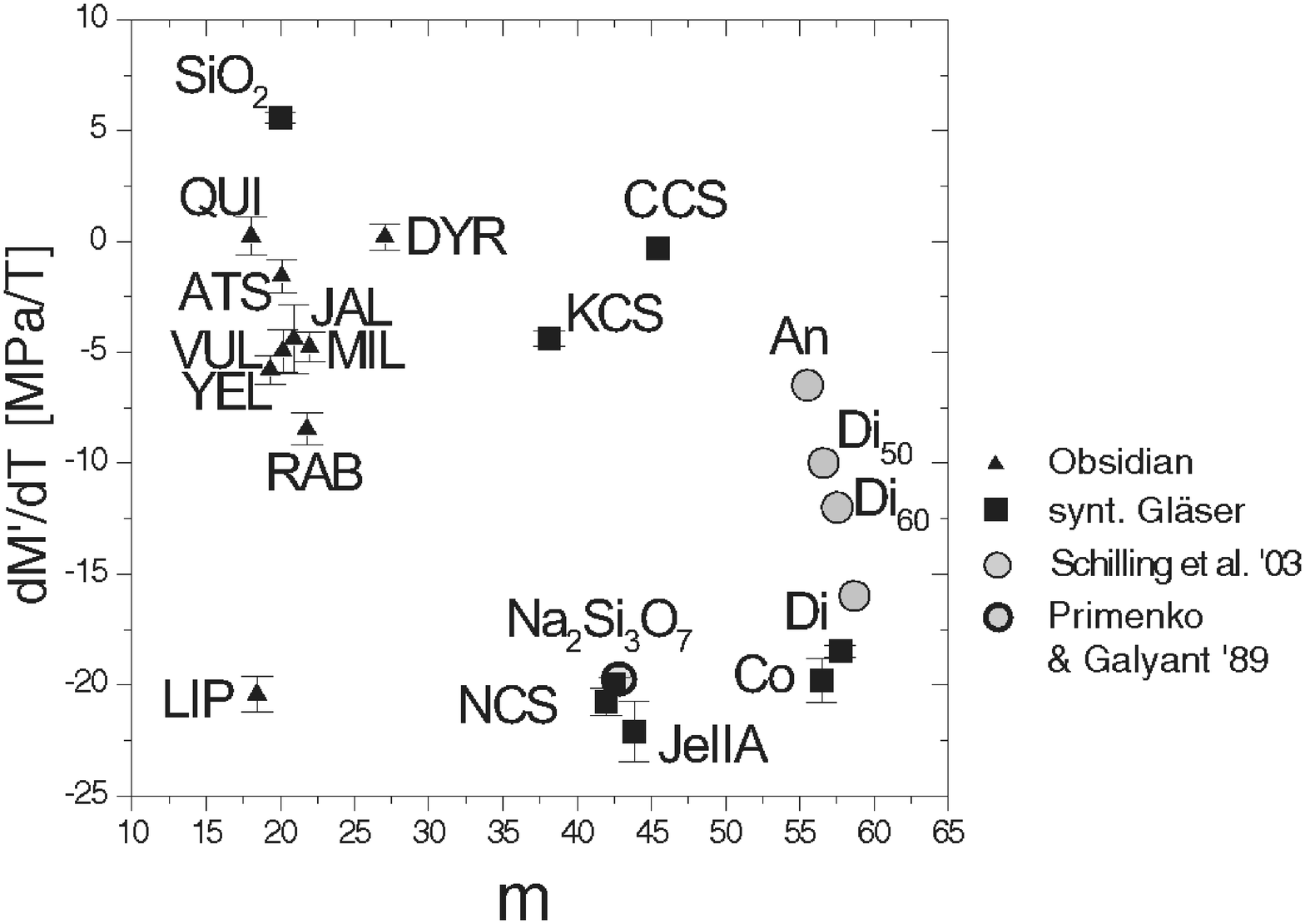}
    \caption{Fragility index $m$ versus the temperature derivative of the storage modulus
             $\partial M'(T)/\partial T|_{T<T_g}$
             in the temperature range below $T_g$ for the obsidians in comparison with synthetic silicate glasses.}
    \label{fig:dMdT}
  \end{center}
\end{figure}

The storage modulus at room temperature $M_{RT}$ is a material constant, i.e. the
Young's-modulus\footnote{If measurements of $M'(T)$ are carried out below $RT$, one observes a
further rise with falling temperature or rising frequency due to relaxation processes
(\cite{Jagd60}).} of the glass $E=M_{RT}$, which ranges for silicate glasses between $60GPa$ and
$100GPa$ (\cite{Wagn04}). The obsidians exhibit a relatively constant value of $(70\pm10)GPa$. In
principle, $E$ is a function of the chemical composition, water-, bubble-, crystal content and
temperature prehistory, however, the sensitivity on these influences is much less than those on
the relaxation times (\cite{Wagn04}). There is a positive correlation with the water content and a
weak negative correlation with the cooling rate (Fig. \ref{fig:WasserModul}).
%%% ------------------------------------
%%% ------------------------------------

In the examined temperature range $M'(T)$ decreases continuously for most silicate glasses (Fig.
\ref{fig:dMdT}). An exception is silica glass with an anomalous increase of Young's modulus with
temperature between $~100K$ and $T_g$ (\cite{Brue70}, \cite{Brue71}). Fig. \ref{fig:dMdT} and Tab.
\ref{tab:ObsidianSpeichermodul} clearly indicate that the examined natural glasses have to be
assigned to the silica glass with the exception of the LIP-obsidian.

\begin{table}[htp]
\caption{Rheological properties of the examined volcanic glasses.  $NaSi$-$Na_2Si_3O_7$;
$^\spadesuit$ \cite{HeidP}, Homosil, $T=1199^\circ$C; $^\diamond$ \cite{Bart96}; $^\P$
\cite{Brue71}; $^\flat$ \cite{Eric75}; $^\natural$ \cite{Poli02};
  $^\#$ \cite{Topl00}; $^\S$\cite{Moch80}, T=RT-350$^\circ$C}
  \label{tab:ObsidianSpeichermodul}
  \begin{tabular}[h]{|l|r|r|r|r|r|r|r|c|}
    \hline
     & $M_{RT}$ & $T_\alpha$ &$M_{\alpha}$ & $-\frac{dM'}{dT}$ & \multicolumn{3}{|c|}{$\frac{\partial\log M'}{\partial (1000/T)}=
                            -\frac{T}{1000}\frac{\partial\log M'}{\partial \log T}$} & $10^{-7}\alpha_T$\\
     & [GPa] & [$^\circ$ C] & [GPa]  & $[\frac{MPa}{K}]$ & $T<T_g$ & $<T<$  & $T_m<T$ & $[1/K]$ \\\hline\hline
   YEL  & 75 & 861& 53 & 5.8 & 0.010 & 0.90 & 16.3 & $\vdots$\\
   VUL  & 77 & 817 & 72 &4.9 & 0.017 & 1.22 & 28.7 & $\vdots$\\
   LIP & 61 & 789&  45 & 20.4 & 0.028 & 0.96 & 14.1 & $\vdots$ \\
   MIL & 67 & 891&  58 & 4.8 & 0.009 & 0.74 & 19.8 & $\vdots$ \\
   RAB & 73 & 819&  55 & 8.5 & 0.025 & 0.9 & 24.4 &  ~61.6$^\flat$\\
   DYR & 72 & 830&  71 & 0.2 & 0.001 & 3.16 & 25.9 & $\vdots$ \\
   ART & 76 & 864&  74 & 1.6 & 0.004 & 1.34 & 18.8 & $\vdots$ \\
   JAL & 70 & 782&  64 & 4.4 & 0.022 & -- & 31,4 & $\vdots$ \\
   QUI & 81 & 824&  81 & -0.2 & -0.001 & 1.47 &  14.4 & \\\hline
   $SiO_2$ &79 & 1239$^\diamond$& ~75$^\spadesuit$ & -5.6 & -0.005 & -0.46$^\natural$ & & 5.4$^\P$ \\
   $Di$  & 92 & 769&  79 & 18.5 & 0.02 & 4.6 & 71 & 139$^\#$ \\
   $NaSi$ & 61 & 523&  55 & 20 & 0.026 & 2.5 & 43.7 & 123$^\S$ \\\hline
  \end{tabular}
\end{table}

\begin{figure}[h]
  \begin{center}
    \includegraphics[scale=0.5]{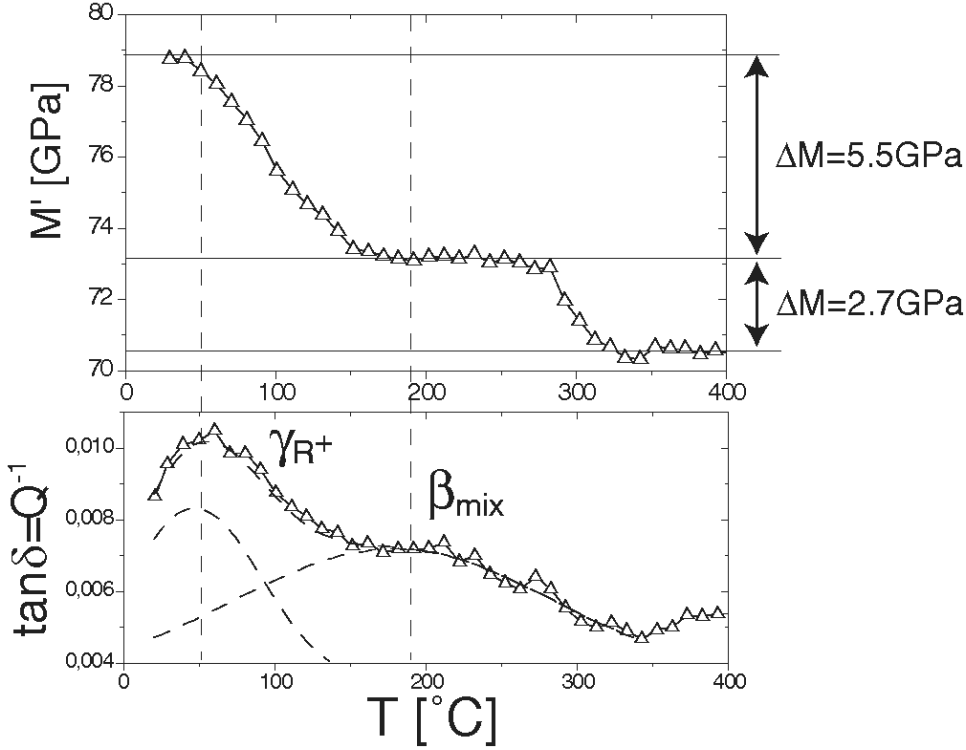}
    \caption{(top) Relaxation steps $\Delta M$ for two relaxation processes: cooperative motion
                of alkali-ions $\gamma_{R^+}$ and mixed alkali-peak $\beta_{mix}$
                 and  (bottom) internal friction $Q^{-1}$ in the case of the $RAB$ obsidian.}
    \label{fig:Stufe}
  \end{center}
\end{figure}

The total temperature dependence of Young's modulus is given by the differential (\cite{Brue71},
\cite{Aska93}, \cite{Rive87}, \cite{VoTh96}, \cite{Schil03}).:
%--------------------------------------
\begin{equation}
    \label{eq:EModulTemperatur}
    \frac{d M'(T)}{d T}=\alpha_T V\left(\frac{\partial M'}{\partial V}\right)_T+\left(\frac{\partial M'}{\partial T}\right)_V
\end{equation}
%--------------------------------------
with volume $V$ and thermal coefficient of expansion $\alpha_T=\frac{1}{V}\frac{\partial
V}{\partial T}$. The sign of $dM'/dT$ will be governed by the sign of $\left(\frac{\partial
M'}{\partial T}\right)_V$, if $\alpha_T$ is negligibly small. In the case of a
Born-von-K$\acute{a}$rm$\acute{a}$n solid, as in the case of a Debye and Grueneisen solid, it was
shown that $dM/dT$ is negative at low temperatures and large $\alpha_T$, and positive at high
temperature and small $\alpha_T$ (\cite{Brue70}). However, steps occur in the storage modulus
curve at the secondary relaxation transitions $\gamma, \beta, \beta'$ (Fig. \ref{fig:Stufe}).
Especially for the natural rhyolitic glasses $d M'(T)/d T$ below $T_g$ is very small and it is
possible to observe the relaxation steps. A constant value of $dM'/dT$ follows at a superposition
of several processes or with a high thermal coefficient of expansion $\alpha_T$ (c.f. and Tab.
\ref{tab:ObsidianSpeichermodul}). If relaxation processes are strong and in different temperature
ranges, then minima occur in $dM'/dT$.
%%% ------------------------------------
%%% ------------------------------------

In the temperature range above $T_g$ $M'$ can be characterized in the Arrhenius-diagram. It can be
differentiated into two ranges: $T_g\leq T\leq T_{\alpha}$ and above $T_{\alpha}$. Here,
$T_{\alpha}$ is the mechanical glass transition temperature at the loss modulus maximum (Tab.
\ref{tab:ObsidianSpeichermodul}). By means of an Arrhenian-ansatz for temperatures $T>T_{\alpha}$
it is possible to determine the apparent activation energy of viscous flow $E'_{a,\alpha}$
(\cite{Wagn04}):
%--------------------------------------
\begin{equation}
   \label{eq:ArrheniusEModul}
   \frac{\partial \log M'}{\partial (1/T)}\approx\frac{2E'_{a,\alpha}}{2,303R}.
\end{equation}
%--------------------------------------

\begin{figure}[p]
  \begin{center}
    \includegraphics[scale=0.57]{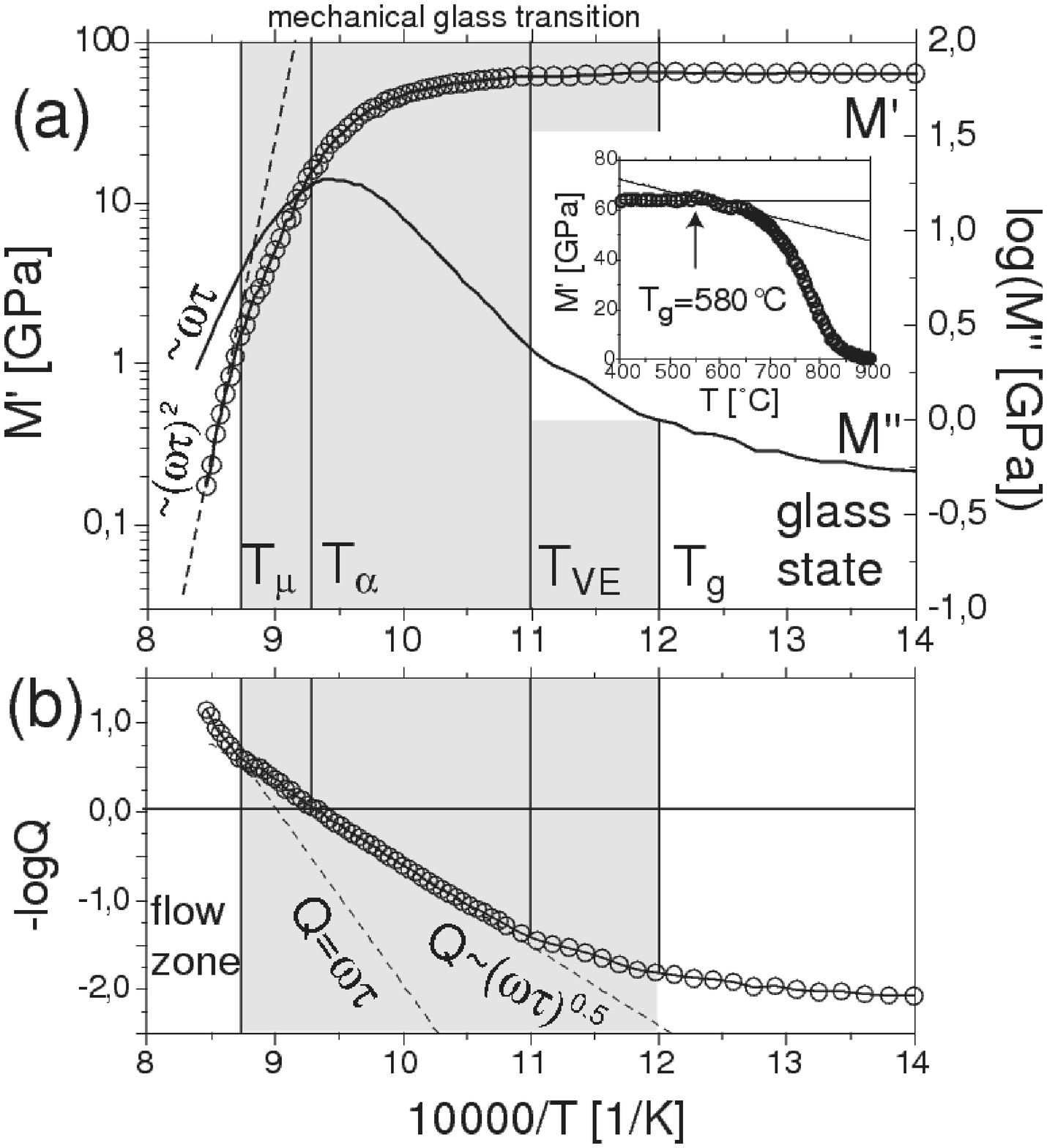}
    \includegraphics[scale=0.4]{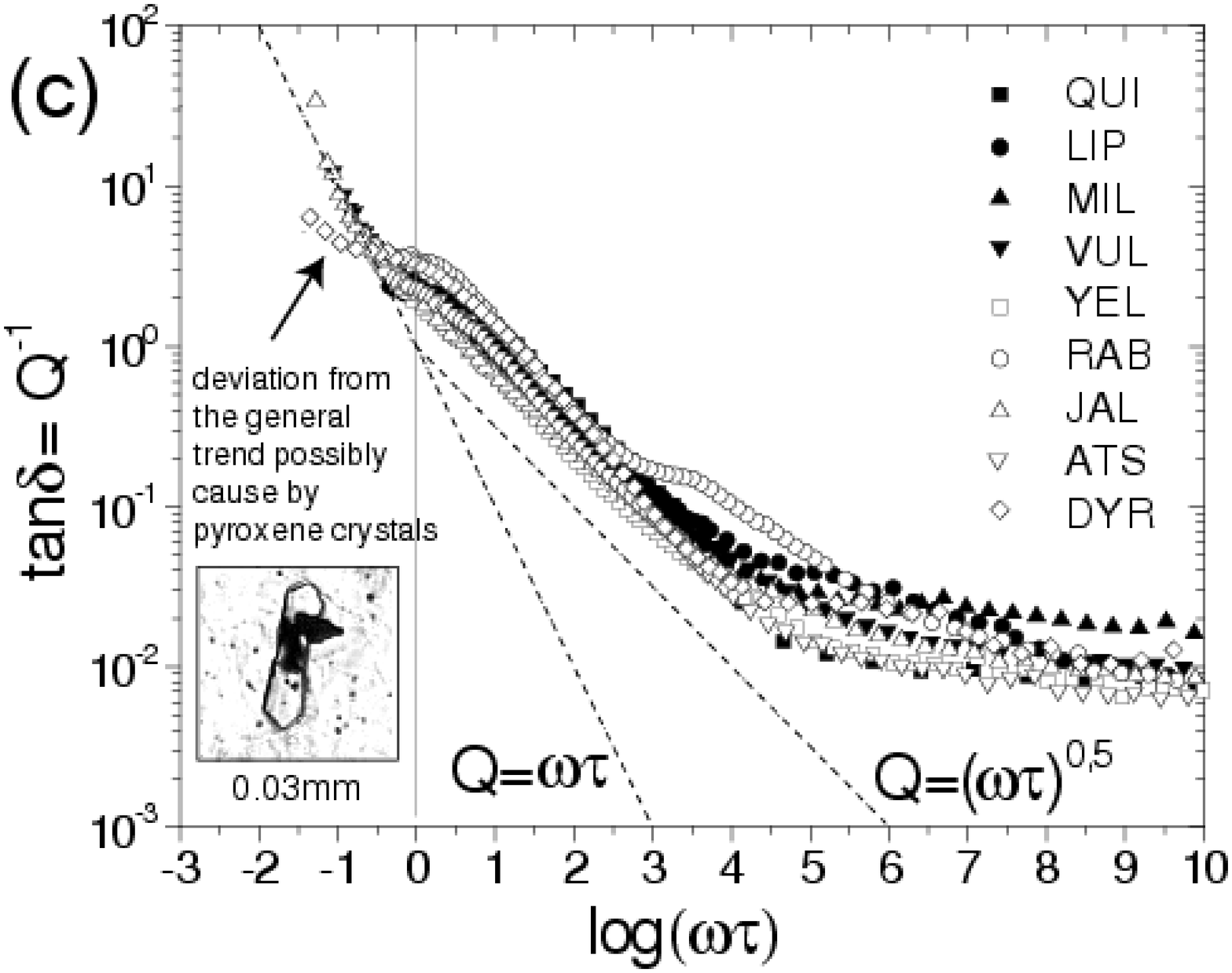}
    \caption{Definition of the temperatures used in this article. (a) Complex Young's modulus $M^\star$ (b) internal
             friction $Q^{-1}$ of the $JAL$-obsidian
             in the Arrhenius diagram. (c) internal friction $Q^{-1}$ as a function of the normalized relaxation time
             $\log(\omega_P\tau_\alpha)$}
    \label{fig:TemperaturenDef}
  \end{center}
\end{figure}

Above $T_g $ all obsidians show a clear temperature dependence of the storage modulus $M'$ which
is caused by the strong change of the structure (configurational states). In principle, it is
possible to determine the glass transition temperature $T_g$ at a relaxation time
$\tau\approx300s$ from the change of $ \partial M'(T)/\partial T$ if a linear decay of $M'$ in the
temperature range below and above $T_g$ is observed. Attention must be paid to distinguish between
$T_g$ at $q\tau\approx1$ (Frenkel-Reiner-Kobeko Gleichung, \cite{Gutz95}), the onset of the
mechanical glass transition $T_{ve}$ at $\omega\tau\approx100$, and the mechanical glass
transition temperature $T_\alpha$ at $\omega\tau\approx1$ (see Fig. \ref{fig:TemperaturenDef}).
%%% ------------------------------------
%%% ------------------------------------

Below $T_{VE}$ no relaxation contributions (viscoelastic relaxation of the structure within the
observation window) can be observed, and it is only possible to examine the mechanical
characteristics of the appropriate ``current" configurational state as well as anelastic
relaxation processes such as the Johari-Goldstein relaxation or the loss due to the mobility of
divalent cations. The temperature dependence is substantially smaller than above $T_\alpha$,
whereby mechanical glass transition is essentially steered by the Maxwell-$\alpha$-relaxation time
$\tau_{\alpha}(T)$.

At relaxed melt viscosity two orders of magnitude above the Maxwell-relaxation time at
$T_\alpha$(\cite {Webb90}, \cite {Webb91}, \cite {Webb92}, \cite {Webb97}) above the temperature
$T_{\mu}\approx930\pm60^\circ C$, should hold the relation
$\lim_{\eta'\rightarrow\eta_{rel}}M'(T)=0$ or
 $\lim_{\eta'\rightarrow\eta_{rel}}G'(T)=0$.

\begin{figure}[h]
  \begin{center}
    \includegraphics[scale=0.7]{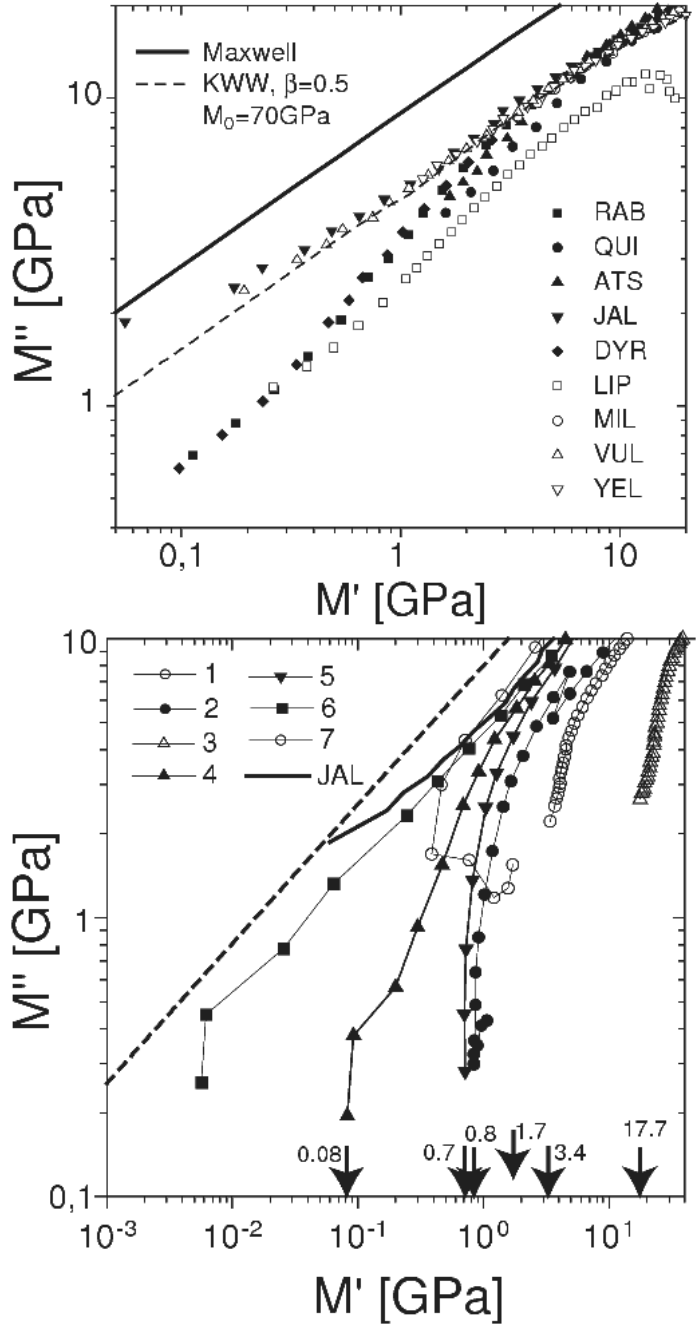}
    \caption{Asymptotic behavior for high temperatures and low frequencies in the representation
                 of the complex modulus $M^\star$ in the Gaussian-plane (top) for the investigated
                 volcanic glasses and (bottom) for starting glass (1: Jenit I, 3: JenitII)
                 and ceramic (2: Jenit I, 4: JenitII) of machinable
                 glass ceramic (\cite{Wagn01}), 5: $16Na_2O-10CaO-74SiO_2$,
                  6: $16Ka_2O-10CaO-74SiO_2$,  7: cordierite-glass which shows surface crystallization effects
                  (\cite{Wagn04}) in comparison to the 8: $JAL$-obsidian.}
    \label{fig:MRestArgand}
  \end{center}
\end{figure}

Since, however, as a function of chemical composition of the glasses phase separation and
crystallization features can arise, and a temperature range of a plateau in Young's and shear
modulus, a relaxed modulus $M'_{rel}, G'_{rel}\neq0$ can occur
 (vgl. Abb \ref{fig:MRestArgand},  \cite{Duff97}, \cite{Duff97a}, \cite{Duff97b},
 \cite{Duff97c}, \cite{Duff98}, \cite{Wagn04}).
%%% ------------------------------------
%%% ------------------------------------
The plateau is typical for Polymers (\cite{Donth92})  in addition, for partially crystalline rocks
or glass ceramics (\cite{Bagd94}, \cite{Bagd97}, \cite{Smit97}, \cite{Lu98}, \cite{Bagd99},
\cite{Bagd00}, \cite{Renn00}). Here, among other things, it was stated that shear stress induced
phase separation can occur (\cite{Qui98}, \cite{Zhan01}, \cite{Arak01}). In the case of obsidians
no plateaus can be observed (Fig. \ref{fig:Module} and \ref{fig:MRestArgand}).

The loss modulus $M''$ of the volcanic glasses is represented in Fig. \ref{fig:Module}. The
mechanical spectrum can be essentially divided into three ranges, characterized by typical
relaxation processes\footnote{The relaxation behaviour in multi-component glasses is rather
complex  due to the superposition of several relaxation processes.}: ($\beta$) low-temperature
range (RT-300$^\circ C$), ($\beta'$) range in the vicinity of $T_g$ and ($\alpha$) viscoelastic,
mechanical glass transition.
%%% ------------------------------------
\begin{figure}[ht]
  \begin{center}
    \includegraphics[scale=0.6]{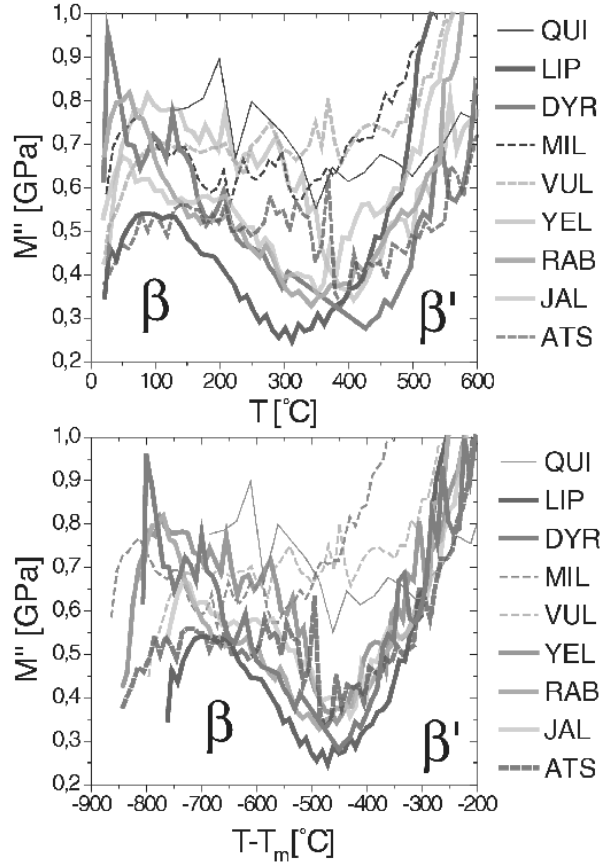}
    \caption{Loss modulus $M''$ as a function of  (top) temperature and (bottom) reduced temperature $T-T_m$
                 in the temperature range $T<T_g$.}
    \label{fig:RohVerlustObsidianeUnterTg}
  \end{center}
\end{figure}
%%% ------------------------------------

Fig. \ref{fig:RohVerlustObsidianeUnterTg} shows the spectra below $T_g$. By scaling the spectra
with $T_\alpha$ it becomes clear that the temperature run of the curve for obsidian of different
sources is very similar. It can be distinguished in two types:  (A) a nearly temperature
independence of the $\beta$-range\footnote{Here it is to be counted for a superposition of several
processes, i.e. a broad $\beta$ transition superimposed by a $\beta'$ process.} of $M''$ with a
average amplitude of $0,7$GPa ($QUI$, $MIL$, $VUL$) and (B) a minimum in $M''(T)$ with
$T-T_\alpha= -450^\circ$ ($LIP$, $DYR$, $YEL$, $RAB$, $JAL$, $ATS$).  The $\beta'$ transition
deviates only for $MIL$-obsidian from the general trend. The temperatures of the minima and
maxima, $T_{\beta, max, min}$ and the values of internal friction are summarized in Tab.
\ref{tab:InnereReibungObsidian}. Since the intensity of the effects and signal-to-noise ratio are
small, all processes below $T_g$ are understood for the $GMM$ as two processes $\beta$ and
$\beta'$.  However, it is in the case of $RAB, JAL, LIP,YEL$ and $ATS$ possible to separate the
$\beta$-process into two overlapping relaxation transitions: cooperative motion of alkali cations
in the glassy network $\gamma_{R^+}$ and a mixed alkali peak $\beta_{mix}$. This overlay was
fitted by means of a double power law function of the form (\cite{Roli98a}):
%%% ------------------------------------
\begin{equation}
\label{eq:Power}
Q^{-1}(T)=Q^{-1}_b+\sum_{j=1}^n\frac{Q^{-1}_{0,j}(\omega\tau_j)^{n_j}}{1+(\omega\tau_j)^{m_j+n_j}}
%\label{eq:lorentz}
%  Q^{-1}(T)=Q^{-1}_0+\frac{2A}{\pi}\frac{w}{4(T-T_m)^2+w^2}
\end{equation}
%%% ------------------------------------
with the background $Q^{-1}_b$, a constant $Q^{-1}_{0,j}$ and the relaxation times
%%% ------------------------------------
\begin{equation}
\label{eq:Power1} \tau_{j}=\tau_{0,j}\exp\left(\frac{E_{a,j}}{RT}\right).
\end{equation}
%%% ------------------------------------
%to the internal friction $Q^{-1}$, with the parameters $Q^{-1}_0, A, w$ and $T_m$.
%%% ------------------------------------
\begin{table}[htp]
  \begin{center}
    \caption{Temperature and magnitude of the internal friction maxima for temperature $T<T_g$ without
                background  subtraction.}
\begin{tabular}{|p{1cm}|r|r|r|r|r|r|r|r|}
\hline
    & $AI$ & $T_{m, \gamma_{R^+}}$ & $Q^{-1}_{m, \gamma_{R^+}}$ &
     $T_{m, \beta_{mix}}$ & $Q^{-1}_{m, \beta_{mix}}$ &
     $T_{min}$ & $Q^{-1}_{min}$ & $Q^{-1}_{Plateau}$ \\
    & & [$^\circ C$] &  & [$^\circ C$] & & [$^\circ C$] & &\\\hline\hline
RAB & 1,06& 54 & 0,010 & 182 & 0,0070 & 342 & 0,005 & --\\
JAL & 0,90 & 63 & 0,009 & 179 & 0,0090 & 342 & 0,006 & -- \\
LIP & 1,08& 91 & 0,009 & 182 & 0,0070 & 293 & 0,004 & -- \\
YEL & 1,07 & 106 & 0,011 & 264 &0,0100 & 395 & 0,005  & -- \\
ATS & 1,29 & 152 & 0,007 & 314 & 0,0070 & 389 & 0,005 & -- \\ \hline\hline
MIL & 1,36& 72 & 0,012 & 142 & 0,0110 & (317) & -- & 0,009 \\
VUL & 1,09 & 121 & 0,009 & 280 & 0,0097 & (355) & -- & 0,0098 \\
QUI & 1,55 & 195 & 0,010 & 431 & 0,0080 & (500) & -- & 0,0076 \\
DYR & 1,46 & -- & -- & 105 & 0,0094 & (431) & -- & 0,004 \\
 \hline
\end{tabular}
    \label{tab:InnereReibungObsidian}
  \end{center}
\end{table}
%%% ------------------------------------
 %%% ------------------------------------
\begin{figure}[p]
  \begin{center}
  \includegraphics[scale=0.5]{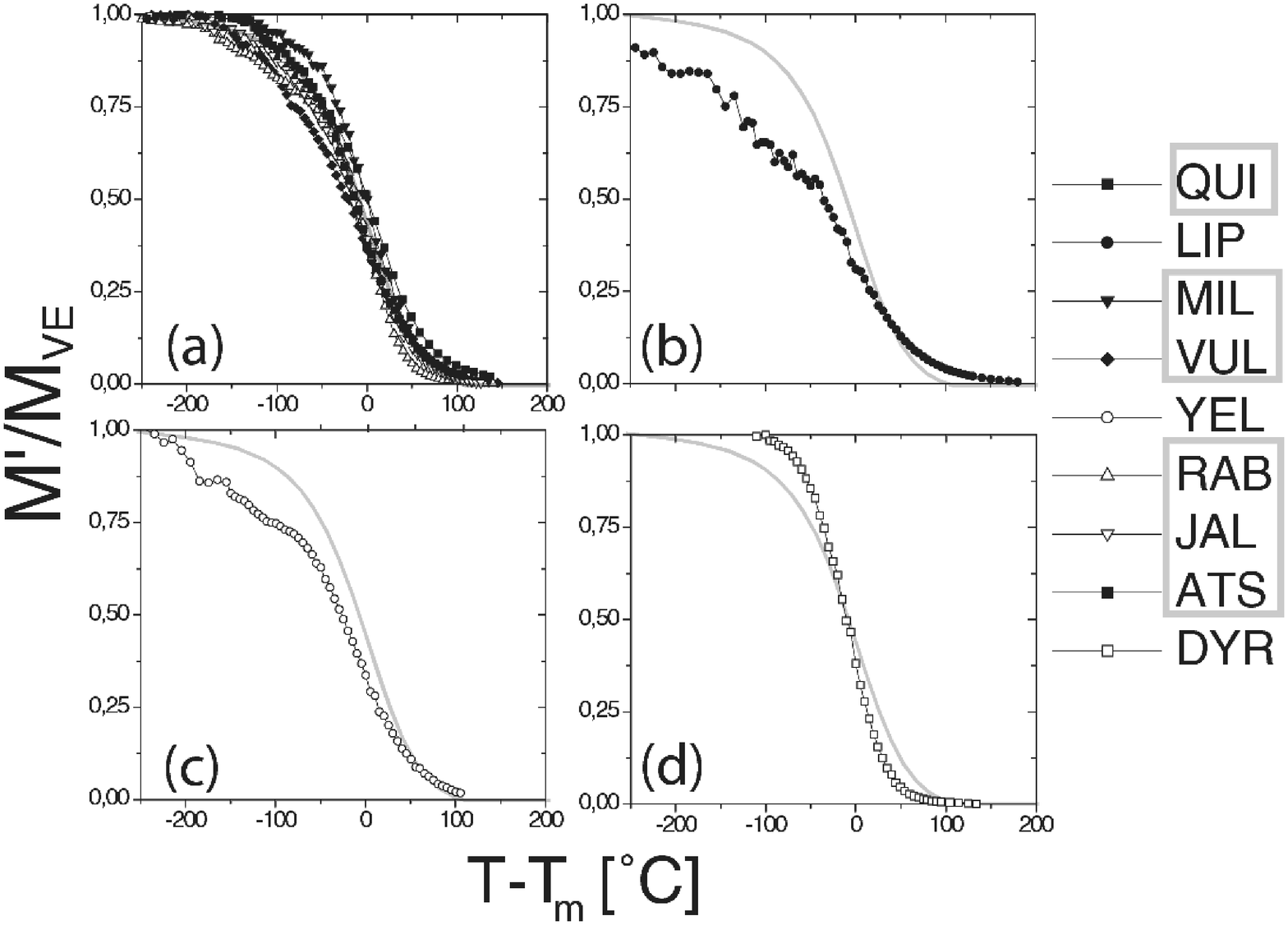}
  \includegraphics[scale=0.5]{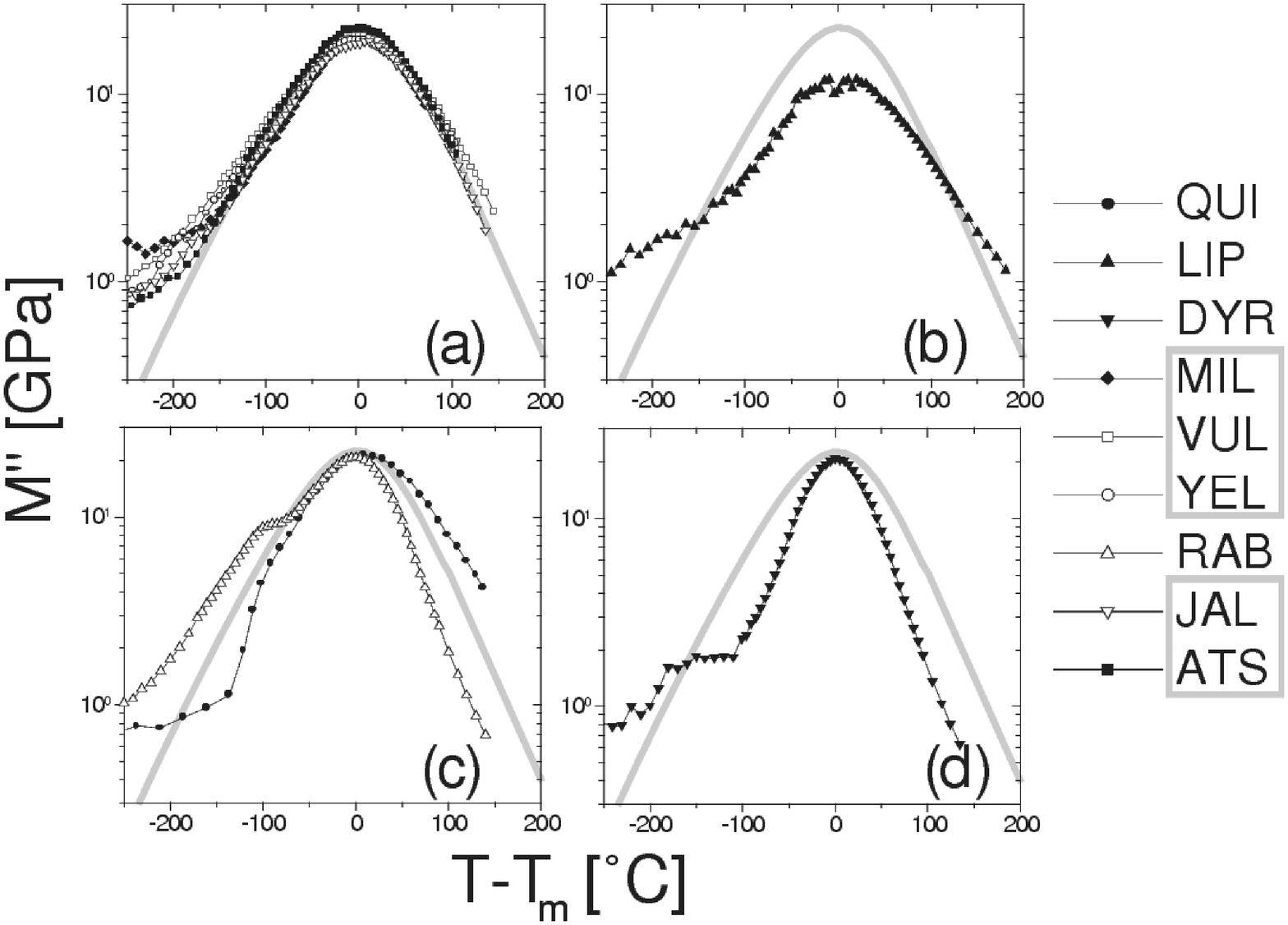}
    \caption{(Top) storage modulus $M'$ normalized with the Young's modulus at the glass transition temperature
     $M_{VE}$ and (bottom) loss modulus $M''$ as a function of $T-T_m$ in the temperature range $T>T_g$.
     In the figures (b)-(d) the deviations from the curve are represented, which the obsidians $MIL$, $VUL$,
     ($YEL$), $JAL$, ($QUI$) and $ATS$ follow and which is clarified with the grey line.}
    \label{fig:KlassifikationVerlustmodul}
  \end{center}
\end{figure}
%%% ------------------------------------

The volcanic glasses can be likewise divided according to the $\alpha$ process into different
groups (Fig. \ref{fig:KlassifikationVerlustmodul}).  By scaling the temperature axes with the
temperature of maxima $T_\alpha$ a master curve for the large group of melts ($MIL$, $VUL$, $YEL$,
$JAL$ and $ATS$) can be constructed, which is an indication for similar rheologic characteristics
of these obsidians.  Deviations from the general trend show the samples $DYR$ and $LIP$. The
$DYR$-obsidian has with $\delta_\alpha=84^\circ C$ a substantially smaller half width and steeper
slopes of the loss modulus flanks of the $\alpha$-maximum:
%%% ------------------------------------
\begin{equation}
\label{eq:Flanken}
  m_i=\frac{d\log M''}{d(10000/T)}, \hspace{0.5cm}i=\left\{ \begin{array}{llr}
                                 1 & \mbox{for} & \log(\omega\tau)<1 \\
                                 2 & \mbox{for} & \log(\omega\tau)>1
                              \end{array}\right.
\end{equation}
%%% ------------------------------------
with $m_1=2.04$, $m_2=-1.56$ than the other glasses (general trend $\delta_\alpha\approx140^\circ
C$ and $1.1<m_1<1.5$, $-1<m_2<-0.6$).  $LIP$ on the other hand exhibits clearly larger
$\delta_\alpha=154^\circ C$ values and a smaller slopes $m_1=1.04$, $m_2=-0.66$.  This obsidian
has the smallest relaxation strength $s_\alpha=12GPa$ and $s_D=27GPa$
($s_\alpha^{average}=20.4GPa$ and $s_D^{average}=35.5GPa$). The $RAB$-obsidian is featured by a
pronounced shoulder\footnote{The shoulder-effect always showed up with all internal friction
measurements accomplished at the obsidian in the same temperature range.}, which suggests a strong
$\beta_{R^{2+}}$ relaxation process. In the $\alpha$ process of the $QUI$-glass an anomaly arises
within the glass transition range. Here, it concerns an physical ageing effect during the
measurement. Particularly, this obsidian exhibits few large ($d\approx0.5mm$) crystals (per sample
approx. 3-4) causing the sample to crack under the experimental strain of $~10^{-4}$.

The mechanical spectra could be parameterisiertparameterized with $GMM$ (\ref{eq:FracMaxwell1})
with the assumption of three relaxation processes ($j=\alpha, \beta, \beta'$):
%----------------------
\begin{equation}
  \label{eq:FracMaxwell2}
   M^*(\omega, \tau)=\sum_{j=1}^3\frac{M_{0,j}M_{1,j}(i\omega\tau_j)^\alpha_j}
         {M_{1,j}(1+(i\omega\tau_j)^{\alpha_j-\beta_j})+M_0(i\omega\tau_j)^\alpha_j}
\end{equation}
%----------------------
\begin{table}[htbp]
\begin{center}\caption{Parameters of the relaxation processes from GMM-fitting (from top to bottom : $\alpha$, $\beta'$ und $\beta$).}
\begin{tabularx}{\linewidth}{|*{7}{X}|}%[h]{|rrrrrrrr|}
\hline Obsidian &   $-\log(\tau [s])$  & $E_a$ & $\beta$ & $M_0$  &  $X$  &  $M_{VE}$ \\
  $\alpha$              &                      &  $[kJ/mol]$     &  & $[GPa]$   &  &  $[GPa]$ \\ \hline
YEL   &    17,02    &    409    &    0,60    &    26    &    0,45    &    59      \\
 VUL    &    17,87    &    384    &    0,53    &    33    &    0,48    &    68    \\
 MIL    &    19,68    &    450    &    0,56    &    60    &    1,00    &    61    \\
 LIP  &    16,13    &    344    &    0,56    &    17    &    0,44    &    39      \\
 DYR    &    24,74    &    554    &    0,65    &    2    &    0,04    &    68     \\
 ATS    &    17,80    &    399    &    0,53    &    49    &    0,66    &    74    \\
 QUI    &    15,74    &    350    &    0,54    &    22    &    0,29    &    77    \\
 ISL    &    19,49    &    442    &    0,70    &    37    &    0,69    &    53    \\
 JAL    &    18,60    &    382    &    0,56    &    45    &    0,78    &    58    \\
\hline
\end{tabularx}
\begin{tabularx}{\linewidth}{|*{7}{X}|}%[h]{|rrrrrrr|}
\hline $\beta'$ &    &      &      &      &      &     \\ \hline
YEL    &    9,17    &    171    &    0,36    &    1,26    &    0,30    &    4,14    \\
 VUL    &    9,28    &    150    &    0,29    &    1,45    &    0,45    &    3,23   \\
 MIL    &    8,61    &    154    &    0,36    &    2,50    &    1,00    &    2,51   \\
 LIP    &    12,39    &    197    &    0,18    &    3,40    &    0,34    &    9,96  \\
 DYR    &    13,92    &    245    &    0,23    &    5,11    &    0,64    &    7,95  \\
 ATS    &    10,48    &    173    &    0,38    &    1,07    &    0,48    &    2,22  \\
 QUI    &    10,80    &    186    &    0,31    &    0,66    &    0,20    &    3,29  \\
 RAB    &    16,82    &    306    &    0,37    &    40,00    &    2,08    &    19,27\\
 JAL    &    11,60    &    190    &    0,31    &    1,40    &    0,39    &    3,58  \\
\hline
\end{tabularx}
\begin{tabularx}{\linewidth}{|*{7}{X}|}%[h]{|rrrrrrr|}
\hline $\beta$ &    &      &      &      &      &     \\ \hline
YEL    &    4,67    &    46    &    0,27    &    1,17    &    0,16    &    7,35     \\
 VUL    &    3,49    &    41    &    0,31    &    0,84    &    0,15    &    5,48    \\
 MIL    &    4,26    &    46    &    0,26    &    0,96    &    0,14    &    6,95    \\
 LIP    &    5,30    &    54    &    0,31    &    1,38    &    0,12    &    11,11   \\
 DYR    &    4,02    &    35    &    0,30    &    0,88    &    0,14    &    6,13    \\
 ATS    &    4,80    &    49    &    0,26    &    1,02    &    0,20    &    4,99    \\
 QUI    &    4,70    &    54    &    0,31    &    0,88    &    0,16    &    5,37    \\
 RAB    &    4,59    &    65    &    0,17    &    0,86    &    0,08    &    11,46   \\
 JAL    &    2,89    &    23    &    0,31    &    1,05    &    0,24    &    4,34    \\
\hline
\end{tabularx}
\label{tab:ParameterAlphaObsidianGMM}
\end{center}
\end{table}

%~~~~~~~~~~~~~~~~~~~~~~~~~~~~~~~~~~~~~~~~~~

%%% ------------------------------------
\begin{figure}[ht]
  \begin{center}
    \includegraphics[scale=0.5]{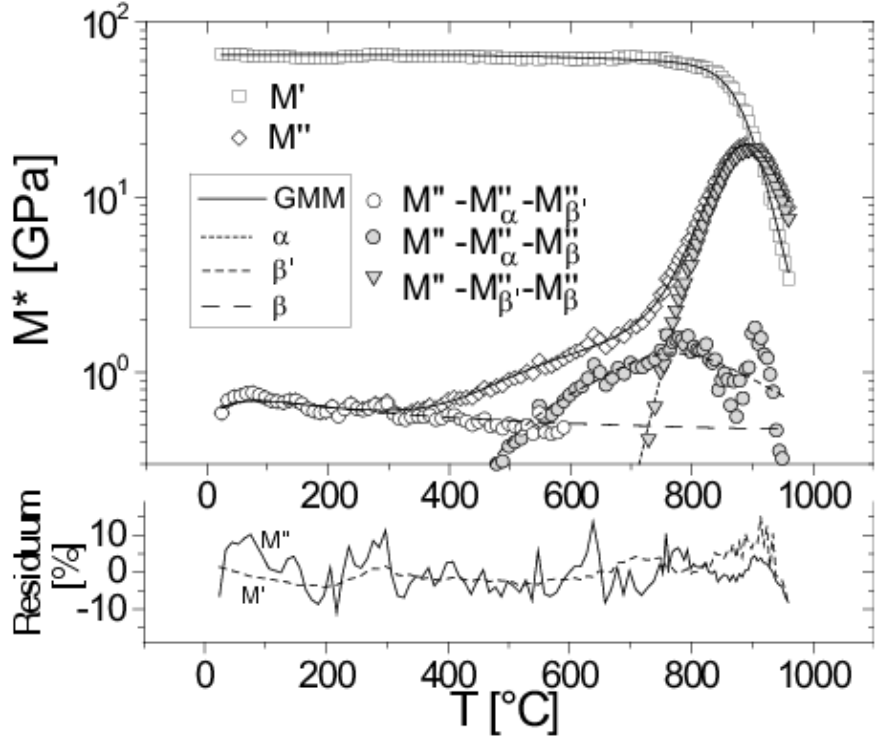}
    \caption{(Top) storage - $M'$ and loss modulus $M''$ as a function of temperature of the $MIL$ obsidian
                with GMM-fit acording to equatio (\ref{eq:FracMaxwell2}).
                 (bottom) Residuum of the fit.}
    \label{fig:AnpassungObsidian1}
  \end{center}
\end{figure}
%%% ------------------------------------
%%% ------------------------------------
\begin{figure}[p]
  \begin{center}
    \includegraphics[scale=0.75]{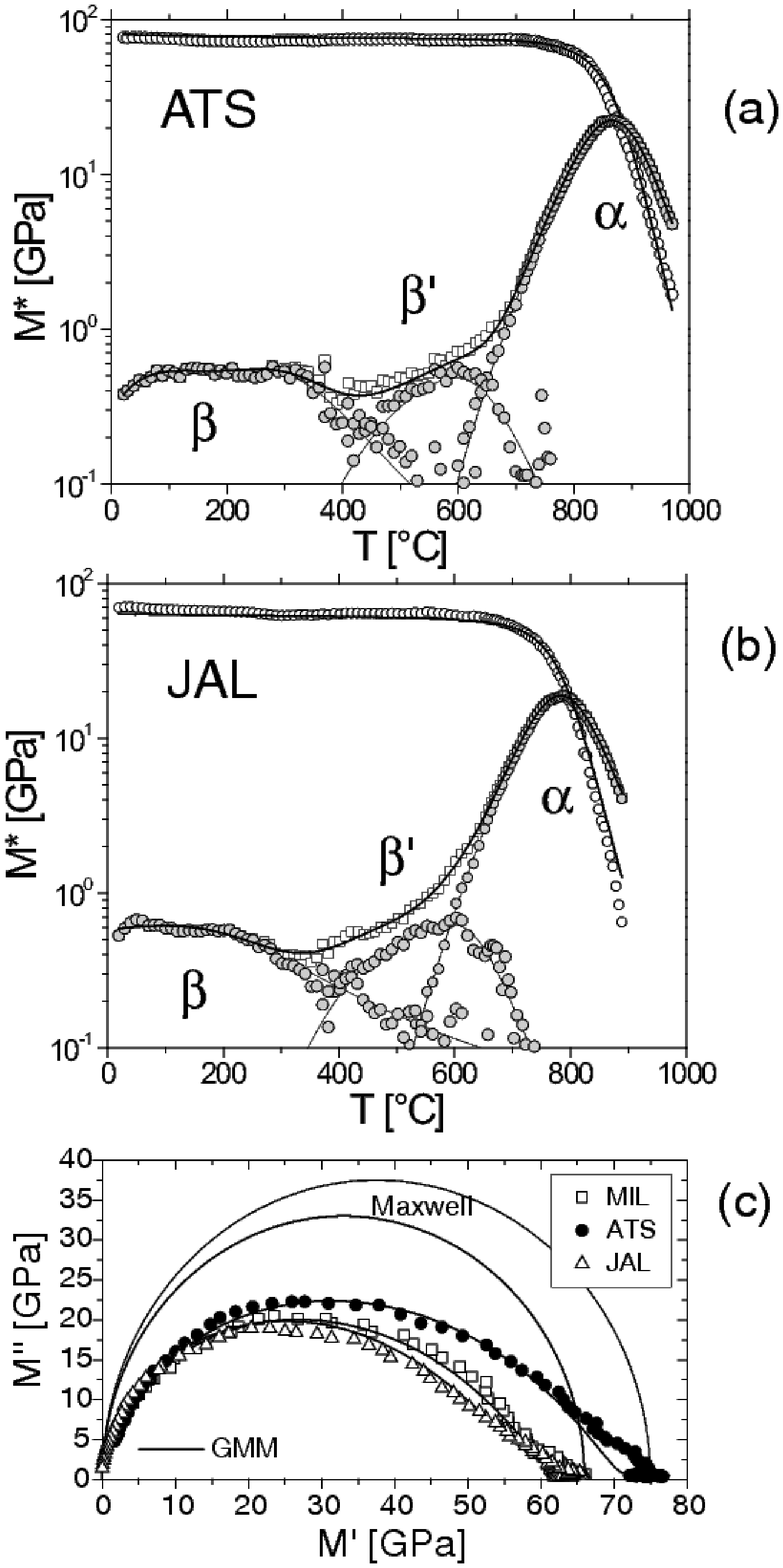}
    \caption{(a) and (b) storage - $M'$ and loss modulus $M''$ as a function of temperature with $GMM$-fit.
                (c) Complex Young's-modulus $M^\star$ with $GMM$-fit represented in the Gaussian-plane
                (selection of three obsidians $MIL$, $ATS$, $JAL$).}
    \label{fig:AnpassungArgand}
  \end{center}
\end{figure}
%%% ------------------------------------

%--------------------------------------%--------------------------------------
\section{Discussion}

Internal friction spectra are shown in Fig. \ref{fig:Module}. Different relaxation processes can
contribute to the relaxation behaviour (\ref{fig:RelaxMapNaSi}): the dynamical glass transition
above the glass transition temperature $T_g$ the so called primary $\alpha$-relaxation
(viscoelastic process) and several secondary relaxation processes below $T_g$: cooperative motion
of alkali ions in the glassy network $\gamma_{R^+}$, mixed alkali peak $\beta_{mix}$, so-called
water peak $\beta_{H_2O}$\footnote{This peak would be present according to \cite{Phal80} and
\cite{Stev85} only in the systems where non-bridging oxygens, $OH^-$ groups and hydrogen bonds are
present at the same time.}, cooperative motion of alkaline earth ions in the viscinity of the
thermal glass transition range $\beta_{R^{2+}}$ (\cite{Roli98}, \cite{Roli01}, \cite{Mart01}) and
a so-called Johari-Goldstein relaxation $\beta_{JG}$ (\cite{Ange00}, \cite{Donth01},
\cite{Nemi03}). It is impossible to distinguish the specific relaxation processes in every case
because of the restricted frequency and temperature range. For this reason three relaxation
processes are assumed: $\beta$ - supperposition of the $\gamma_{R^+}$, $\beta_{mix}$ and
$\beta_{H_2O}$ process, $\beta'$ - superposition of the $\beta_{R^{2+}}$, $\beta_{JG}$ and the
thermal glass transition as well as the viscoelastic $\alpha$-transition.

Further it is assumed, that the unrelaxed modulus $M_\infty$ is a very weak function of
temperature. This is true in the case of volcanic glasses (Fig. \ref{fig:dMdT} and Tab.
\ref{tab:ObsidianSpeichermodul}).

Since volcanic glasses are natural materials, these are heterogeneous up to certain degrees. The
natural glasses have a smaller signal-to-noise ratio cause by these heterogeneities opposite to
synthetic glasses. Thus, sensitivity is limited. Relaxation processes can only seriously be
observed in relaxation behaviour if they are strongly enough. The different assumed influences and
their effect on different relaxation processes are summarized in Tab. \ref{tab:UebersichtEffekte}.
On one hand it is shown, that there are processes, which are sensitive for different influences,
c.f. $\beta'$. On the other hand, the $\alpha$ process is caused by the chemical composition and
water content. In addition, effects arise due to degassing and vesiculation. Here, however, it
showed up clearly that even during very strong vesiculation of the $LIP$-sample, dynamics of the
glass transition is detectable. In this specific case, temperature dependence of
$\alpha$-relaxation time can only be modelled in first approximation by an Arrhenian-equation due
to water release.
%----------------------%----------------------
\begin{table}[htbp]
  \begin{center}
    \caption{Assumed influences on the mechanical relaxation behaviour of the volcanic glasses:
                 X - significant, (X) - weak, (-) - uncertain}
    \begin{tabular}[h]{|l|p{3cm}ccc|}
      \hline
      influence & $\beta$-process & $\beta'$-process & $\alpha$-process & $M_{RT}$\\
      \hline\hline
        chemical composition & X & (-) & X & (X) \\
        water content            & (X) & (-) & X & (X) \\
        bubble content (degassing)    & (X) & (X) & (X) & (X) \\
        crystals                    & (X) & (X) & (-) & (-) \\
        temperature history    & X & (X) & (-) & (-) \\\hline
     \end{tabular}
    \label{tab:UebersichtEffekte}
  \end{center}
\end{table}
%----------------------%----------------------

%----------------------%----------------------
\subsection{Relaxation behaviour below $T_g$}
%----------------------%----------------------

The influences of the chemical composition and the water content on the relaxation processes below
$T_g$ were examined in detail in several mechanical relaxation studies ($Q^{-1}$-spectroscopy) of
synthetic glasses. It was shown that different mixing cation effects exist, which in addition
depend on the water content and polymerization degree of glasses (c.t. \cite{Day74a},
\cite{Phal80} and \cite{Stev85}, \cite{Mart01}). The examined volcanic glasses exhibit the
following ions:  $Na^+, K^+, Ca^{2+ }, Mg^{2+ }, Fe^{2+}$ and water in different speciation
(\cite{Gabe99}, \cite{Dorm00}, \cite{Behr01}, \cite{Dore02}), which can contribute to the
mechanical relaxation behaviour.  Since the place-exchange processes can be observed both in the
internal friction spectrum $Q^{-1}(T)$  or in equivalence in the loss modulus $M''(T)$ as maxima
and in the storage modulus  $M'(T)$  as a step. The following views are accomplished on the basis
of internal friction.

The $JAL$-obsidian is peralkalin, i.e. it exist on $NBO$ bounded alkali ions $R^+$ ($\approx
1$mol\%), which are mobile in the glass network.  The peak of $\gamma_{Na^+}$ at a measuring
frequency of $~1Hz$ is to be expected below $RT$. Investigations of internal friction of binary
$R_2O$-$SiO_2$-glasses with $R^+ = Rb^+, K^+, Na^+, Li^+$ show (\cite{Zdan76}) that the intensity
$Q_{m, \gamma}^{-1}$ and the temperature $T_{\gamma_{R^+}}$ of this maximum depends on the field
strength $F=\frac{z}{a^2}$ of appropriate cations with the valence $z$ and the Van der Waals
radius $a$.  It becomes clear that the $\gamma_{R^+}$ process is coupled with the $\beta_{H_2O}$
process, which exhibits a maximum at approximately $200^\circ C$.

If one  further regards ternary glasses with two different kinds of alkali ions, then one observes
the occurrence of a further maximum: the mixed alkali peak $\beta_{mix}$ within the range of the
water maximum. At a temperature of  $\approx320^\circ C$ additional relaxation processes arise: a
$\beta'_{R^{2+}}$-process in direct connection to the mobility of alkaline-earth ions, the
mobility  of oxygen-ions with an activation energy of $\approx200kJ/mol$ and the Johari-Goldstein
relaxation (\cite{Roli01}, \cite{Nemi03}, \cite{Wagn04}). A substantial conclusion from the
investigations of \cite{Mart01} on ternary alkali and alkaline-earth silicate glasses is the fact
that in systems with two kinds of cation, while one is substantially faster than the other one,
the cation radius relationship affects the activation energy of the slower ions. The more similar
the ion radii are, the smaller is the activation energy.

By the component $Al_2O_3$ the location and intensity of the maxima change drastically. For the
ternary sodium alumosilicate glass examined by \cite{Day62} as well as \cite{Day62a} the
$\gamma_{Na^+}$ as well as $\beta_{H_2O}$ peak decreases with rising $Al_2O_3$-content, whereby at
the ratio $Al_2O_3/Na_2O=1$ the $\gamma_{Na^+}$ process is maximal and the $\beta_{H_2O}$ process
cannot be observed anymore.

Measurements of mixed alkali alumosilicate glasses with $Al_2O_3/R_2O=1$ of \cite{Saka85}as well
as \cite{Saka89} point out, that as in the case of the ternary mix alkali glasses, these systems
exhibit a mixed alkali maximum $\beta_{mix}$ at $\approx100^\circ C$. At a ratio $Na_2O/K_2O=1$
arise a very strong maximum, which is formed by the superposition of the $\gamma_{R^+}$ and
$\beta_{mix}$ processes.

\cite{Saka85} perform additional measurements of glasses, in which the $Al_2O_3$-content is varied
at the expense of  the $SiO_2$-content, at constant $Na_2O/K_2O=19$.  By the presence of $Na^+$
and $K^+$-cations is formed again apart from the $\gamma_{R^+}$ a mix alkali maximum
$\beta_{mix}$. With rising $Al_2O_3$ the intensity of the $\gamma_{R^+}$ peak increases, while
those of the $\beta_{mix}$ decreases. As in the case of ternary sodium alumosilicate glasses from
the investigation of \cite{Day62} as well as \cite{Day62a} acquiring the intensity of the
$\gamma_{R^+}$ maximum and of $\beta_{mix}$ minimum at $Al_2O_3/Alk=1$.

From the reasons specified above, the following consequences resulted in the mechanical relaxation
behaviour below $T_g$ of the examined natural glasses:
\begin{itemize}
  \item The water maximum is suppressed strongly by the $Al_2O_3$-content.
  \item The ratio $Al_2O_3/Alk\approx1$ leads to the formation of an intensive $\gamma_{R^+}$ peak.
  \item The presence of two different alkali ions $K^+$ and $Na^+$ leads to the formation of a mixed alkali peak
          $\beta_{mix}$ with $T_{\beta_{mix}}\approx140^\circ C$.
  \item The $\gamma_{R^+}$ process is superimposed by the $\beta_{mix}$ process
          with the result of a relatively high internal friction $Q^{-1}\approx0.7$
          within the temperature range $RT-300^\circ C$.
  \item By the presence of divalent cations a further relaxation process $\beta_{R^{2+}}$
        arises in the vicinity of the glass transition range.
\end{itemize}
The compilation of data to the internal friction of volcanic glasses confirms the above
assumptions (Fig. \ref{fig:RohVerlustObsidianeUnterTg}).  In the case of the samples $LIP, RAB,
YEL$ and $JAL$ it is suggested that it concerns a superposition of two processes.  $ATS, MIL, QUI$
and $VUL$ exhibit a pronounced plateau.  An exception is the $DYR$ sample. Here it seems that the
$\gamma_{R^+}$ peak is below $RT$. It is suggested a further peak occurs at $\approx100^\circ C$.
In Tab. \ref{tab:InnereReibungObsidian} the results are summarized.
%%% ------------------------------------
%\begin{figure}[htp]
%  \begin{center}
%    \includegraphics[scale=0.5]{InnereReibungObsidian.eps}
%    \caption{Internal friction $Q^{-1}$ as a function of the temperaturer $T$ in the temperature range $T<T_g$.}
%    \label{fig:InnereReibungObsidian}
%  \end{center}
%\end{figure}
%%% ------------------------------------
As mentioned, the decay of the storage modulus with temperature below $T_g$ is likewise dependent
on the chemical composition.  In the special case of the obsidians it can be assumed that $M_{RT}$
depends secondarily on the chemical composition, due to ion mobility in the glassy network, but is
determined essentially by water content.  However, practically no investigations of the influence
of water content on elastic properties of synthetic and natural glasses exist in contrast to
internal friction measurements. The occurrence of the $LIP$ sample is interesting in this context
with the smallest Young's modulus of $M_{RT}=61GPa$. It is possible to calculate $M_{RT}$ with the
model of \cite{Priv01} (SciGlass information system) in accordance with the measured data.

%----------------------%----------------------
\subsection{Primary $\alpha$-Process}
%----------------------%----------------------
\subsubsection*{a) Chemical composition and water content}
%----------------------%----------------------
%%% ------------------------------------
\begin{figure}[t]
  \begin{center}
    \includegraphics[scale=0.5]{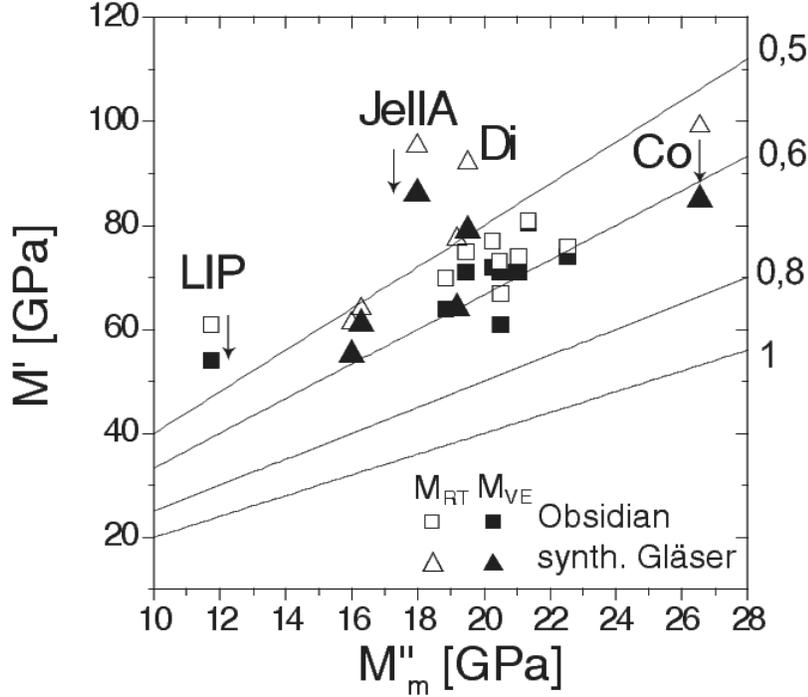}
    \caption{Comparison of the Young's modulus at RT $M_{RT}$ or $T_g$ $M_{VE}$ and loss modulus
     at the $\alpha$ peak
             $M''_{\alpha,m}$ of the obsidians and several sythetic glasses (\cite{Wagn04}).}
    \label{fig:M1M2Glaeser}
  \end{center}
\end{figure}
%%% ------------------------------------
The comparison of the Young's modulus at RT ($M_{RT}$) or $T_g$ ($M_{VE}$) and loss modulus at the
$\alpha$ peak ($M''_{\alpha,m}$) is represented in Fig. \ref{fig:M1M2Glaeser}. The drawn in lines
refer to the deviation from simple Maxwell behavior $(x=1)$:  $M''_m=xM'_{VE, RT}/2$. The linear
regression results are $x_{RT}=0,55\pm0,02$ and $x_{VE}=0,58\pm0,02$. The flat curve of the $LIP$
sample is to be explained by strong vesiculation. A change of the sample caused by degassing and
formation of bubbles leads to a change in the mechanical behavior, whereby the viscoelastic
response is dominated by the mechanical glass transition. The parameter $x$ is a measure
relatively independent of the chemical composition, with which deviations can be interpreted as
physically induced processes in contrast to the influence of the chemical composition or the water
content, which causes a change of the properties of the glass.
%%% ------------------------------------
\begin{figure}[ht]
  \begin{center}
    \includegraphics[scale=0.5]{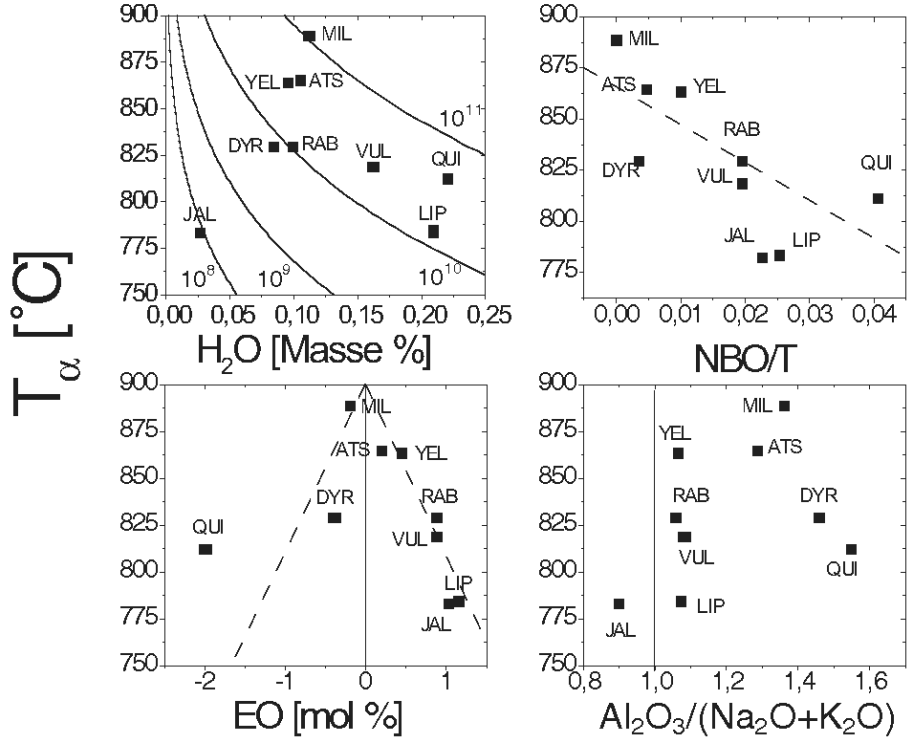}
    \caption{Temperature of the $\alpha$-peak $T_{m,\alpha}$ as a function of the chemical composition
                 and water content of the investigated volcanic glasses. The \cite{Hess96} calculation scheme defined by the solid
                 curve for viscosities ($10^8$...$10^{11}$Pas) closely fits the experimental data for the calc- and per-alkaline
compostions.
                 The dashed lines are an aid to the eye, and do not conform to any specific mathematical expression.}
    \label{fig:TmObsidianeChemie}
  \end{center}
\end{figure}
%%% ------------------------------------
The $GMM$-fitting supplies the unrelaxed modulus at the mechanical glass transition $M_{VE}|_ {
T_{VE}}$ which is $\approx2GPa$ lower than the elastic modulus at $T_g$.  In the used model the
$\beta'$ process represents an overlay of a secondary\footnote{In the case of the obsidians the
mobility of divalent cations and the Johari-Goldstein relaxation.} process and the thermal glass
transition caused on experimental conditions. This secondary effect is observed likewise in
organic glasses with dielectric spectroscopy, and gave reason to controversial discussions of the
existence of the Johari-Goldstein-process (\cite{Nemi03}, \cite{Schn00}, \cite{Ange00}).

The occurrence of the $\alpha$-process is defined by the chemical composition and the water
content (\cite{Ding95}, \cite{Hess96b}, \cite{Stev98}, \cite{Gior03}). In Fig.
\ref{fig:TmObsidianeChemie} one can notice different trends.  The temperature $T_\alpha$ decreases
with increasing $NBO/T$ and the amount of the absolute excess oxide value ($|EO|$), i.e. with
increasing depolymerization. Likewise the glass transition temperature as well as the temperature
of water release follow this trend. The mechanical glass transition of the $MIL$ obsidian with the
smallest $NBO/T=2^{-4}$ and a $|EO|$ close to zero $\approx0,17$ occurs at the highest
temperature.  The smallest maximum temperature $T_{\alpha, \alpha}$  has the peralkaline $JAl$
obsidian with $NBO/T=0,023$ and $|EO|=1,05$.  Even the $LIP$ obsidian follows this trend. The
occurrence of the mechanical glass transition (thus) is dominated by the chemical composition (see
also the investigations of partially crystalline glasses by \cite{Muel03}).  This conclusion
confirm measurements of Cordierite-glass (\cite{Wagn01}, \cite{Wagn04}).  Impurities (crystals,
bubbles, cracks) influence the width of the maxima and relaxation strength.
%%% ------------------------------------
\begin{figure}[htp]
  \begin{center}
    \includegraphics[scale=0.5]{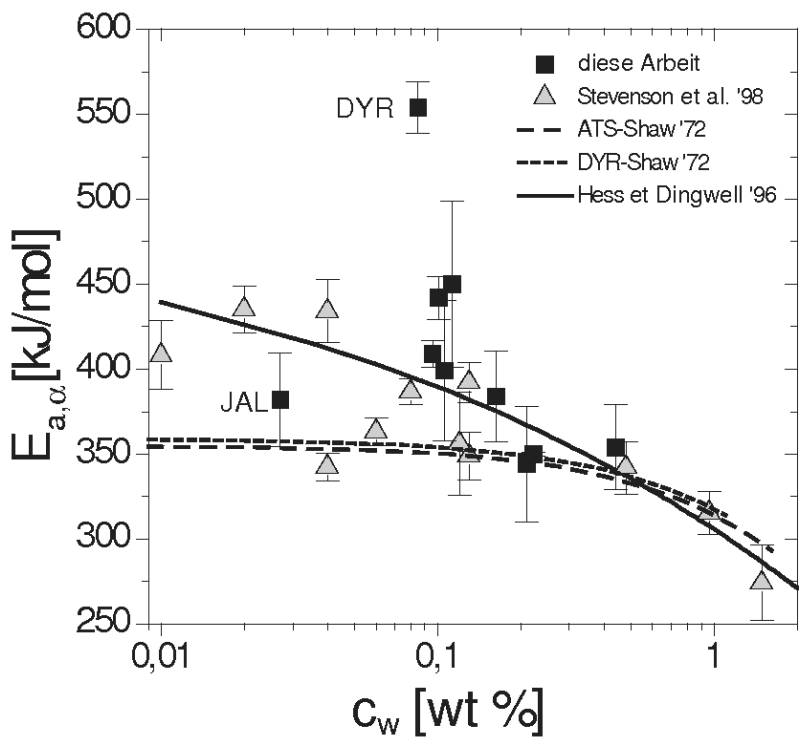}
    \caption{Activation energy of viscouse flow $E_{a, \alpha}$ at the glass transition temperature
             $T_g$ as a function of water content for the natural volcanic glasses of this work and
             the natural and remelted obsidians of \cite{Stev98}.}
    \label{fig:AI-DatenAutoren}
  \end{center}
\end{figure}
%%% ------------------------------------
The activation energy and prefactor of the Arrhenius-equation for the temperature dependence of
the relaxation time $\tau_\alpha(T)$ can be determined from the $GMM$-fit and compared with the
results of the measurements of other authors (viscosity measurements, torsion pendulum
measurements) (Fig. \ref{fig:AlSiDatenAutoren }). The activation energy of the $\alpha$ relaxation
time or viscosity at $T_g$ is essentially a function of chemical composition and water content. A
model that considers both influences sufficient for synthetic or natural silicate melts does not
exist. The model of \cite{Priv01} represents the most progressive at present, in the connection
between structure and properties (volatile excluded).  For the description of volcanic glass, the
empirical model of \cite{Hess96a} is most effective. This model was developed for water bearing
leucogranitic melts but was also used to calculate the viscosity of metaluminous and peraluminous
obsidian (\cite{Stev98}).
%%% ------------------------------------
%\begin{figure}[htp]
%  \begin{center}
%    \includegraphics[scale=0.4]{AlSiDatenAutoren.eps}
%    \caption{Activation energy of viscouse flow $E_a$ at the glass transition temperature $T_g$
%                 as a function of the molar ratio $Al/(Al+Si)$
%                 for difference natural ans synthetic silicate melts.}
%    \label{fig:AlSiDatenAutoren}
%    \includegraphics[scale=0.4]{AI-DatenAutoren.eps}
%    \caption{$E_a$ as a function of the agpaitic index $AI=(Na_2O+K_2O)/Al_2O_3$ (mol \%).}
%    \label{fig:AI-DatenAutoren}
%  \end{center}
%\end{figure}

%%% ------------------------------------
%
%\paragraph{In summary} the following points has result:
%\begin{itemize}
%  \item The mechanical $\alpha$-relaxation (mechanical or viscoelastic glass transition)
%          superpose a secondary $\beta$-process (Abb. \ref{fig:MasterAlleDaten}).
%  \item The occurance and the width at half high is defined by the temperature dependence of the viscosity.
%  \item The  width of the glass transition $\delta$ is therefor a function of the fragility  and the distribution of relaxation times.
%  \item The relaxation time distribution is in comparison to the fragility index for anorganic glasses nearly constant
%          (Abb. \ref{fig:betaFD-Ea-alleDaten}-\ref{fig:MasterAlleDaten}).
%  \item The presence of crystalls and bubbles $>1$-$2 Vol.\%$ alter the dynamic of the glass transition.
%  \item The chemical composition and the volatile content define the fragility index.
%\end{itemize}

%%% ------------------------------------
\begin{figure}[htp]
  \begin{center}
    \includegraphics[scale=0.5]{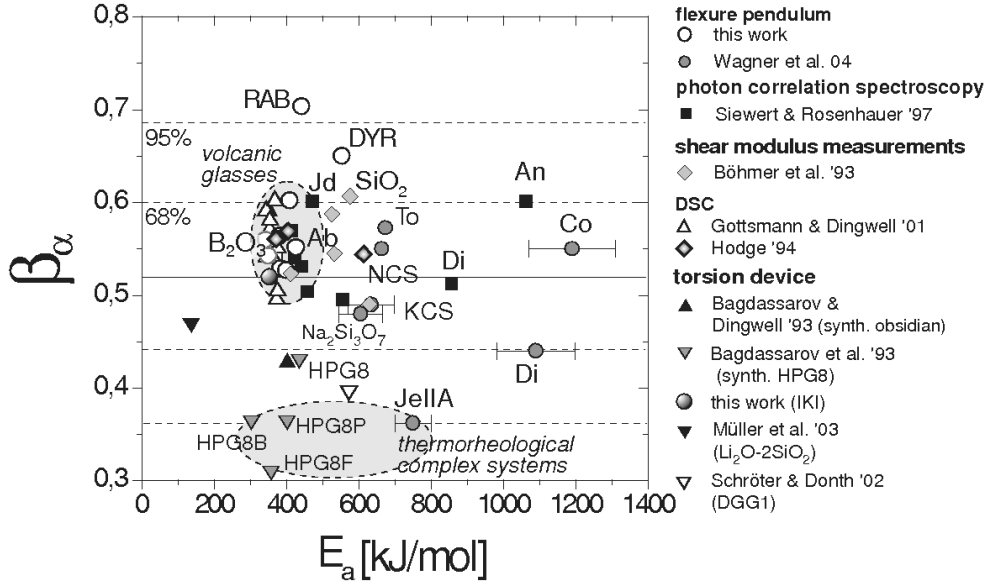}
    \caption{Distribution parameter $\beta_a$ ($\alpha_\alpha=1$) as a function of the apparent activation
    energy for viscous flow $E_{a, \alpha}$ at $T_g$ of the synthetic and volcanic melts of this work
    in comparison to literature data. The  $\beta_\alpha$ parameter was calculated from $\beta_{KWW}$
    with an empirical expression.}
    \label{fig:betaFD-Ea-alleDaten}
  \end{center}
\end{figure}
%%% ------------------------------------

Fig. \ref{fig:betaFD-Ea-alleDaten} show the range of variation of the distribution of relaxation
times on the basis of the $\beta_{\alpha}$ parameter and activation energy $E_a$ at the mechanical
glass transition temperature for structurally different silicate melts. The
$\beta_{\alpha}$-parameter for the obsidians range between $0.5$ and $0.6$ with two exceptions
$DYR$ and $RAB$. The $RAB$ glass has a very strong $\beta'$ relaxation process so that the
GMM-fitting leads to a narrower $\alpha$-process. In comparison, the $DYR$ glass shows anomalous
behaviour in all examined parameters.

The average value of all supercooled melts is $\overline{\beta}_{\alpha}=0,52$. Deviations can be
observed for two volcanic glasses as well as the HPG8-melts from the work of \cite{Bagd93a}.
%%% ------------------------------------
\begin{figure}[htp]
  \begin{center}
    \includegraphics[scale=0.5]{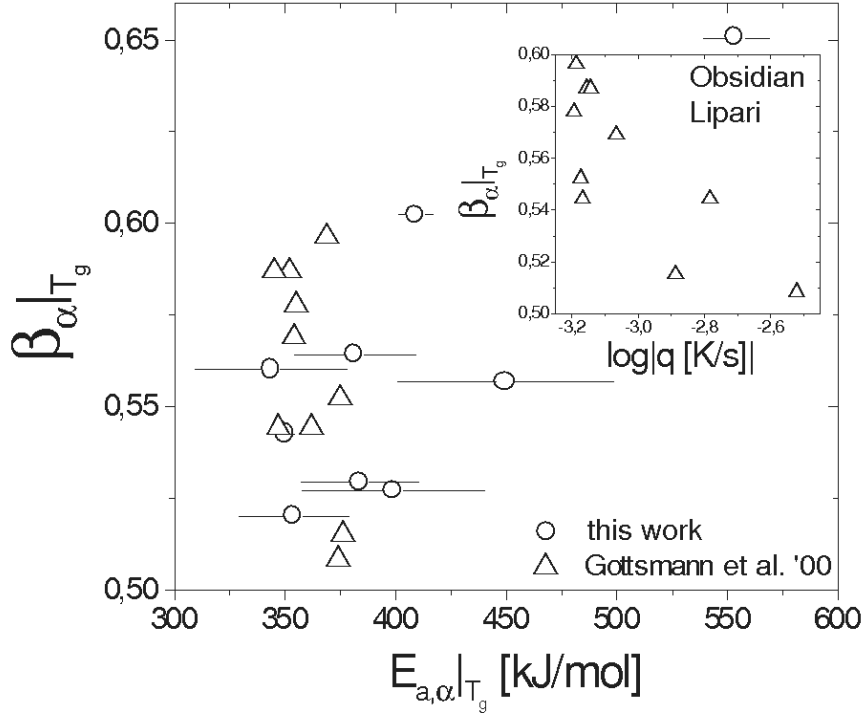}
    \caption{Distribution parameter $\beta_a$ ($\alpha_\alpha=1$) as a function of the apparent activation
    energy for viscous flow $E_{a, \alpha}$ at $T_g$ of the volcanic melts. (inset) Distribution parameter $\beta_a$
    ($\alpha_\alpha=1$) as a function of the cooling rate $|q|$ of a $LIP$-obsidian (\cite{Gott01}).}
    \label{fig:AI-DatenAutoren}
  \end{center}
\end{figure}
%%% ------------------------------------

Under the condition that the distribution of relaxation times is nearly constant it should be
possible to predict the occurrence and the width of the mechanical glass transition from viscosity
data (see \cite{Wagn04c}).

The variation of the distribution parameter is primarily caused by phase separation and
crystallization processes as well as degassing and vesiculation. The temperature prehistory has
little influence. The $DSC$ investigation of \cite{Gott01} shows a negative correlation of the
distribution parameter and the cooling rate. The variation of the $\beta$ parameter is within the
standard deviation of the 45 samples.

\paragraph*{In summary} the following points result:
\begin{itemize}
  \item The mechanical $\alpha$ relaxation (mechanical or viscoelastic glass transition)
         overlays a secondary $\beta$ relaxation process.
  \item The occurrence of the transition is defined by the temperature dependence of viscosity.
  \item The width $\delta$ of the transition depends on the fragility and the distribution of relaxation times.
  \item The distribution of relaxation times in the investigated temperature and frequency range is almost constant
          in comparison to the fragility index for inorganic glasses (see \cite{Wagn04}).
  \item The presence of crystals (in particular due to surface crystallisation) and bubbles $>1$-$2 Vol.\%$ changes the
         dynamics of the transition and therefore the relaxation time distribution.
  \item Fragility is defined by the chemical composition and the presence of volatiles.
  \item Phase separation and degassing processes change strongly the dynamics and thermics
         \footnote{Due to the nomenclature of \cite{Dont81} and \cite{Donth01}.}
        of the transition and lead to therorheologic complex behavior.
 \end{itemize}

\section{Conclusion}

Obsian can considered as a thermorheological simple melt with a low fragility index $m\approx20$.
Variations in the ratio $NBO/T$ determine the polymerization degree of the melts and thus the
rheologic properties. A crucial factor which determines the mechanical relaxation behaviour in the
thermal and dynamic glass transition range above $T_g$, is the relatively high water content. In
the vitreous state the influence of water on the dynamical processes plays a minor role, since the
volcanic glasses are almost completely polymerized ($NBO/T\leq0.4$). For clarifying this fact it
requires more exact methods.

\paragraph*{Three section}can be distinguish in the mechanical spectra (a) $T<T_g$, (b) $T\approx
T_g$ und (c) $T>T_g$, with the following characteristics:
\newcounter{Fakt}
\begin{list}{(\textbf{\roman{Fakt}})}{\usecounter{Fakt}}
  \item The Young's modulus $M_{RT}=(70\pm10)$GPa is nearly constant. There is a positive correlation with the water content
         and a weak negative correlation with the cooling rate (\cite{Wagn04}).
  \item Below $T_g$ the mechanical relaxation behaviour is characterized by the cooperative motion of univalent cations
          (alkali-peak $\gamma_{R^+}$, mixed alkali-peak  $\beta_{mix}$) and the polymerisation degree.
           Impurities (bubbles, crystals, cracks, trace elements) in the natural glasses decrease the signal to noise ratio.
  \item All glasses exhibit a very weak, and in one case, an anomalous negative decay of the Young's modulus with temperature like
          silica glasses.  In addition, it is possible to observe steps in the curve of Young's modulus ($\Delta M'<5$GPa)
          which correspond with the  $\gamma_{R^+}$ or $\beta_{mix}$ peak in internal friction.
  \item In the vicinity of the glass transition temperature there is a superposition of the $\alpha$ and a $\beta'$ process
          (cooperative motion of divalent cations: alkali earth peak $\beta_{R^{2+}}$ and the Johari-Goldstein
           relaxation $\beta_{JG}$) as well as the thermal glass transition defined by the experimental conditions.
          The thermal glass transition is reflected clearly in the loss modulus and in the storage modulus at least less than $1\%$.
  \item The occurence of the mechanical or dynamic glass transition  at a constant frequency as function of the temperature is specified by
          the temperature dependence of the structural $\alpha$ relaxation time $\tau_{\alpha}$.
 \item The temperature dependence of the structural $\alpha$ relaxations time is determined by the polymerization
         degree with water as a quasi controlling parameter and can be computed by means of an
         empirical equation according to \cite{Hess96a} or \cite{Hess96b}
          for the per- and metaluminous rhyolitic melts.
         The model of \cite{Priv98} as well as \cite{Gior02} for the computation of the temperature dependence of the viscosity
         of multicomponent melts cannot be used for the computation of the $\alpha$ relaxation time.
         For strong melts and melts with high  water content the glass transformation temperature
         and the activation energy of viscous flow are overestimated.
 \item Degassing processes in connection with bubble growth as well as other impurities (crystals)
         lead to macroscopic (cracks) and structural
         (polymerization degrees) changes, which cause instabilities of the samples.
 \item The dynamics of the mechanical glass transition is a universal process with very week variation of the relaxation time distribution
         almost independent of chemical composition and water content (0,027... 0.44)Masse$\%$ of the examined
         natural melts.
\end{list}

The obsidians represents an interesting material class, which  has many useful properties
(concerning the methods of the mechanical spectroscopy)\footnote{"`Impurities"' have to keep in
mind, which are able to effect the actual glass characteristics, like in the case of the
$LIP$-sample}, like thermalrheologically simple behavior, small fragility index and a high glass
formation tendency, small variation of the Young's modulus in a wide temperature range, high
polymerization degree, long-term resistance.
%\end{linenumbers}

%----------------------------------------------------
\bibliography{Literatur}
\bibliographystyle{elsart-harv}%
%----------------------------------------------------
\end{document}